\documentclass[twocolumn,twocolappendix]{openjournal}

\usepackage{xcolor}
\usepackage{textgreek}
\usepackage[utf8]{inputenc}
\usepackage[english]{babel}
\usepackage{amsmath} 
\usepackage{bm}

\usepackage{hyperref}
\hypersetup{ unicode, colorlinks=true, linkcolor=linkcolor, citecolor=linkcolor, filecolor=linkcolor, urlcolor=linkcolor, }
\usepackage{color,colortbl}
\definecolor{linkcolor}{rgb}{0.0,0.3,0.5}
\usepackage{tensind}
\tensordelimiter{?}
\DeclareGraphicsExtensions{.bmp,.png,.jpg,.pdf}
\usepackage{verbatim}
\usepackage[normalem]{ulem}
\usepackage{orcidlink}
\usepackage{soul}
\usepackage{url}
\newcommand{\unitspace}{\,}
\newcommand{\cm}{\ensuremath{\unitspace \mathrm{cm}}}
\renewcommand{\sec}{\ensuremath{\unitspace \mathrm{s}}}
\newcommand{\Msun}{\ensuremath{\unitspace \mathrm{M}_{\odot}}}

\renewcommand{\vec}[1]{ \bm{#1}} 
\newcommand{\nvec}[1]{ \hat{\bm{#1}}} 
\newcommand{\ud}{\mathrm{d}} 

\newcommand{\lamC}{\ensuremath{\lambdabar_\mathrm{C}}}

\begin{document}

\title{Pair Discharges and Radio Emission from Pulsar Magnetospheres}

\author{Joonas N\"attil\"a\orcidlink{0000-0002-3226-4575}}
\email{joonas.nattila@helsinki.fi}
\affiliation{Department of Physics, University of Helsinki, P.O. Box 64, FI-00014 University of Helsinki, Finland}

\author{Tuomo Salmi\orcidlink{0000-0001-6356-125X}}
\email{tuomo.salmi@helsinki.fi}
\affiliation{Department of Physics, University of Helsinki, P.O. Box 64, FI-00014 University of Helsinki, Finland}

\begin{abstract}
Radio pulsars can power their coherent radio emission through intermittent plasma discharges in magnetospheric gaps.
The nonlinear coupling between particle acceleration, quantum electrodynamic (QED) processes, and electric field screening remains difficult to model self-consistently.
We present an analytical concurrency model and first-principles particle-in-cell (PIC) simulations of polar cap discharges.
The model predicts limit-cycle behavior where the electric field and plasma density oscillate with a frequency set by the local plasma frequency, modified by the plasma inertia.
One-dimensional simulations with exact QED rates and realistic plasma parameters validate these predictions.
The discharges generate pair multiplicities up to $M_\pm \sim 10^4$ and excite electric field oscillations with spectra consistent with pulsar radio microstructure.
These results provide a theoretical framework connecting microphysical gap dynamics to macroscopic observables.
\end{abstract}


\maketitle

\section{Introduction}

Pulsar radio emission originates from the star's magnetosphere \citep[see][for a review]{philippov2022}.
A typical radio pulsar rotates with an angular frequency $\Omega_\star \equiv 2\pi/P$, where the period $P \sim 1 \sec$.
The magnetic field lines co-rotate with the star up to the light cylinder radius $R_\mathrm{LC} \equiv c/\Omega_\star$.
This radius marks a transition zone where the magnetic field topology changes from predominantly closed to open \citep{goldreich1969, sturrock1971}.
In the closed zone, the magnetic field $\vec{B}$ co-rotates with the star and both magnetic foot-points connect to the star's surface.
In the open zone, the $\vec{B}$-field twists and only one foot-point attaches to the surface.
Global magnetospheric models describe this structure in detail \citep{countopolous1999,gruzinov2005,spitkovsky2006, timokhin2006,kalapotharakos2012}.
We focus on the open field-line region.

The rotating magnetosphere requires a charge density of $\eta_\mathrm{co} = - \vec{\Omega} \cdot \vec{B}/(2\pi c)$ to enforce co-rotation by canceling the rotationally induced electric field component parallel to $\vec{B}$ \citep{goldreich1969}.
The flow of this plasma can sustain a maximum (field-aligned) co-rotation current density $j_\mathrm{co} = c \eta_\mathrm{co}$.
In comparison, the twist in the open field-line bundle induces a current $\vec{j}_\mathrm{m} = (c/4\pi) \vec{\nabla} \times \vec{B}$ along the field lines that connect the polar regions and the edge of the light cylinder.
This current $j_\mathrm{m}$ can over-supply the required current ($\alpha \equiv j_\mathrm{m}/j_\mathrm{co} > 1$) or provide it with the wrong sign ($\alpha < 0$), leading to intermittent regions with unscreened parallel electric fields, known as gaps \citep{shibata1997,beloborodov2008,timokhin2010}.
Here we focus on the polar gap above the magnetic poles of the star \citep{ruderman1975}.
Other gaps include, e.g., the slot gap \citep{arons1983}, the outer gap \citep{cheng1986}, and return-current gap(s) \citep{bransgrove2023}.

Plasma discharges intermittently screen the gaps \citep{timokhin2010, timokhin2013}.
Quantum electrodynamic (QED) processes enable electron-positron pairs to interact with high-energy photons, populating the gap with copious plasma clouds of high pair multiplicity $n_\pm e / |\eta_\mathrm{co}| \gg 1$ \citep[e.g.,][]{daugherty1982,arendt2002,hibschman2001,timokhin2019}, where $n_\pm$ is the local pair number density.
Previous work developed analytical models for discharge dynamics \citep{cruz2022,tolman2022, okawa2024} and simulated QED cascades with radiative particle-in-cell (PIC) \citep{timokhin2010, timokhin2013, philippov2020, cruz2021, chernoglazov2024, benacek2024, benacek2025} and Vlasov methods \citep{ye2025}.

Here we present a new analytical concurrency model that simultaneously describes the coupled dynamics of plasma multiplicity, photon production, and electric field evolution.
The discharge exhibits a stable limit-cycle behavior driven by the interplay between particle acceleration and screening.
Comparison with one-dimensional radiative PIC simulations including exact QED processes validates the theoretical model and characterizes the resulting particle spectra and electromagnetic radiation.

\newpage

\section{Pulsar Magnetospheres}\label{sect:2}

\subsection{System Parameters}

Radio pulsars are neutron stars with a radius $R_\star \approx 10^6 \cm$ and mass of $M_\star \approx 1.4 \Msun$ \citep[e.g.,][]{nattila2022c}.
They rotate with a period $P \sim 1 \sec$, corresponding to an angular velocity $\Omega_\star \approx 6 \,P_0^{-1} \,\mathrm{s}^{-1}$.
We use $Q_x \equiv Q/10^x$ to denote the value of a quantity $Q$ in units of $10^x$ (cgs).

A typical radio pulsar has a magnetic field strength of $B_\star \sim 10^{12}\,\mathrm{G}$ at the surface.
This value approaches the Schwinger limit $B_Q \equiv m_e^2 c^3 / (\hbar e) \approx 4.4 \times 10^{13}\,\mathrm{G}$, so that
\begin{equation}
  b \equiv \frac{B_{\star}}{B_Q}
    \sim 0.02 \,B_{12} \, ,
\end{equation}
rendering QED mildly nonlinear.
Far from the star, the magnetic field structure resembles a dipole, whereas close to the surface it can have higher multipolar components with deformations and magnetic loops similar to those of the Sun.

The rotating magnetosphere splits into open and closed zones (see Fig.~\ref{fig:circuit}).
The light cylinder radius, where azimuthal velocity $v_\phi = c$ for the rotating field, is $R_\mathrm{LC}/R_\star \sim 5000 \,P_0$.
The separatrix angle between the last closed field line and the dipole axis $\vec{\mu}$ at $r = R_\star$ is $\sin\theta_\mathrm{pc} \approx \sqrt{ R_\star/R_\mathrm{LC} }$, giving a polar cap size
\begin{equation}
  R_\mathrm{pc} \approx R_\star \sin\theta_\mathrm{pc}
                \approx R_\star \left( \frac{R_\star}{R_\mathrm{LC}} \right)^{1/2}
                \sim 0.01  \,P_0^{-1/2} ~R_\star \, .
\end{equation}
A curvature radius for a dipole field is $R_\mathrm{curv} \approx 4 r/(3\sin\theta)$ (where $r$ is the distance from the star and $\theta$ is the colatitude angle from the dipole axis);
at the edge of the polar cap it is $R_\mathrm{curv} \approx \frac{4}{3}\sqrt{R_\mathrm{LC} R_\star} \sim 9 \times 10^7 \,P_0^{1/2} \cm$.
We take, instead, $R_\mathrm{curv} \sim R_\star \sim 10^6 \cm$ to account for a possible multipole structure.

Assuming that the magnetic axis and spin axis are aligned, the polar cap's rim rotates with a velocity
\begin{equation}
  \beta_\mathrm{rot} \approx \frac{\Omega_\star R_\mathrm{pc}}{c}
                     \approx 3 \times 10^{-6}  ~P_0^{-3/2} \, ,
\end{equation}
twisting the open field-line tube from the stellar surface.
The rotational velocity $\vec{v}$ of the moving magnetic field induces an electric field $\vec{E} = -\vec{v} \times \vec{B}/c$.
The electric field at the polar cap surface is poloidal,
\begin{equation}
  E_\theta \sim - \frac{\Omega_\star R_\mathrm{pc} B_r}{c}
           \approx - \beta_\mathrm{rot} B_\star \, ,
\end{equation}
which defines the characteristic rotational field scale $E_\mathrm{rot} = \beta_\mathrm{rot} B_\star$ used below.
The electric field also determines the electrostatic potential difference (where $\ell$ is a spatial coordinate along the polar cap radius)%
\footnote{%
In Cartesian coordinates with $\vec{B} = B \nvec{z}$ and $\vec{v} = \beta c x \nvec{y}/ L$, we have $\vec{E} = -\vec{v} \times \vec{B}/c = -\beta  B x \nvec{x}/L$.
The potential drop is then $\Delta V = \int_0^L \beta B (x/L) \ud x= \beta B L/2$.
    \label{footnote:cartesian}
} $
    \Delta V =
    - \int \vec{E} \cdot \ud \vec{\ell}
    \sim \beta_\mathrm{rot} R_\mathrm{pc} B_\star/2
   \sim 7 \times 10^{12} \,B_{12} P_0^{-2} \,\mathrm{V}$.
When the gap above the polar cap cannot supply the charge density required to maintain $E_\theta$, the electric field adjusts so that $\vec{\nabla}\!\cdot\vec{E}\neq 0$, forcing a radial component to develop.
Near the pole, where $B_r \gg B_\theta$, this radial component becomes the parallel accelerating field $E_\parallel$.

\subsection{Global Circuit}

\begin{figure}[t!]
\centering
\includegraphics[clip, trim=0.0cm 0.0cm 30.0cm 6.0cm, width=0.99\columnwidth]{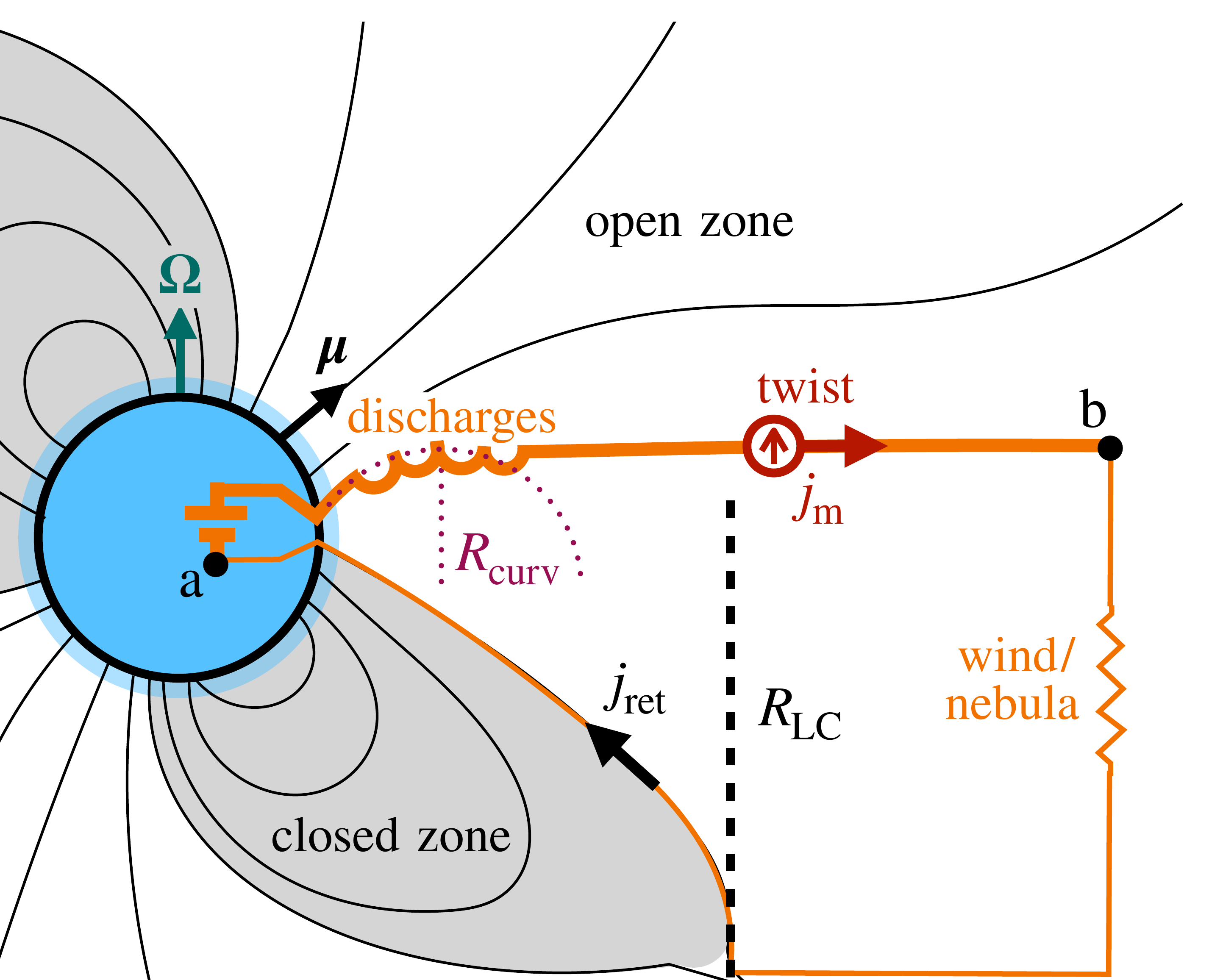}
\caption{\label{fig:circuit} Schematic visualization of the pulsar magnetosphere (thin black curves), the studied electric circuit (thick orange curve), and the relevant circuit components (see text). We simulate the one-dimensional circuit from point a to b. }
\end{figure}

The rotating magnetosphere forms an electric circuit with two competing elements (see Fig.~\ref{fig:circuit}):
rotational electric field and twist current of the field lines.
These two sources do not balance in general, resulting in magnetospheric regions with unscreened (parallel) electric fields, known as gaps.

First, rotation of a magnetized conductor induces a surface charge distribution that supports the co-rotation electric field.
This distribution determines the sign of charges extracted from the polar caps.%
\footnote{ Aligned rotators extract electrons from both polar caps, while anti-aligned rotators extract positive charges (ions or positrons).}
The resulting charge separation sustains the co-rotation (Goldreich--Julian) charge density, of magnitude $|\eta_\mathrm{co}| = e n_\mathrm{co}$, where%
\footnote{%
In Cartesian coordinates as in Footnote~\ref{footnote:cartesian}, we have $| \vec{\nabla} \cdot \vec{E} | =  \beta B/L$ and $n_\mathrm{co} = \beta B/(4\pi e L)$.
}
\begin{equation}\label{eq:ngj}
  n_{\mathrm{co}} = \frac{|\vec{\nabla} \cdot \vec{E}|}{4 \pi e}
    \sim \frac{\Omega_\star B_\mathrm{d}}{2 \pi c e}  \approx %
    \frac{\beta_\mathrm{rot} B_\star }{4 \pi e R_\mathrm{pc}}
    \sim 10^{10} \,B_{12}\,P_0^{-1} \cm^{-3} \, ,
\end{equation}
where $B_\mathrm{d} = B_\star/2$ is a magnetic field strength characterizing the dipolar field.
The co-rotation density can also be expressed through the plasma frequency, $\omega_\mathrm{p,co} = \sqrt{4\pi e^2 n_\mathrm{co}/m_e}$, which satisfies the relation $\omega_\mathrm{p,co}^2 = \Omega_\star \omega_B$, where $\omega_B \equiv e B_\star/(m_e c)$ and $\Omega_\star = 2\pi/P \approx \beta_\mathrm{rot} c/R_\mathrm{pc}$.
The corresponding skin depth is $c/\omega_{\mathrm{p,co}} \sim 3 \cm$, which sets the characteristic scale over which electromagnetic fields are screened in the plasma.
The plasma frequency satisfies $\omega_\mathrm{p,co} = \sqrt{\sigma_\mathrm{co}} \Omega_\star$, where the co-rotation magnetization parameter
\begin{equation}
  \sigma_{\mathrm{co}} = \frac{B_\star^2}{4\pi n_\mathrm{co} m_e c^2}
                       \sim 10^{18} \,B_{12} \, P_0\, ,
\end{equation}
confirms that the magnetospheric plasma at the co-rotation density $n_\mathrm{co}$ is highly magnetized.

Second, the twisted open field-line tube carries an electric current $\vec{j}_\mathrm{m} = (c/4\pi) \vec{\nabla} \times \vec{B}$.
Its relative strength \citep{mestel1985,shibata1997, beloborodov2008, timokhin2013}
\begin{equation}
  \alpha \equiv \frac{j_\mathrm{m}}{j_\mathrm{co}}
\end{equation}
compared to the maximum current sustained by the co-rotation density $j_\mathrm{co} \equiv \pm e n_\mathrm{co} c$ is determined by global magnetospheric dynamics.

The parallel electric field along the $\vec{B}$-field line (with a spatial coordinate $h$;
$h=0$ marks the surface) in the co-rotating frame evolves according to (see Appendix~\ref{app:corot}):
\begin{align}
  \partial_h E_\parallel &\approx 4\pi (\eta - \eta_\mathrm{co} ) \, , \\
  \partial_t E_\parallel &\approx 4\pi (j_\mathrm{m} - j_\pm) \, .
\end{align}
The field vanishes, $E_\parallel(h, t) = 0$, only if $\eta = \eta_\mathrm{co}$ and $j_\mathrm{m} = j_\pm = \eta c \beta_\parallel =\eta_\mathrm{co} c \beta_\parallel=j_\mathrm{co} \beta_\parallel$, where $\eta$ is the local charge density, $j_\pm$ the field-aligned current, and $\beta_\parallel$ the particle velocity along the $\vec{B}$-field;
we also assume $E_\parallel=0$ at the boundary.
This implies that $|\alpha| < 1$, because $|\beta_\parallel| < 1$.
The real plasma flow in the circuit is more complex because particles at the surface are at rest ($\beta_\parallel(h=0) = 0$), leading to spatial variations in $\beta_\parallel$.
The flow can be categorized into three regimes depending on the strength and sign of $\alpha$:

i) \textbf{Sub-corotating flow.} For $0 < \alpha < 1$, the time-stationary solution exhibits spatially oscillatory motion with momenta varying between $p_\parallel \in [0, p_\mathrm{max}]$, where $p_\mathrm{max} = 2\alpha/(1 - \alpha^2)$ \citep{beloborodov2008}.
While the stationary solution is purely spatially oscillatory, time-dependent kinetic simulations (with free particle supply from the star) show that it is unstable:
phase-space mixing develops and the flow evolves into a chaotic trapped-particle state \citep{timokhin2013}.
For typical pulsar parameters, $p_\mathrm{max}$ is insufficient to ignite pair cascades.

ii) \textbf{Super-corotating flow.} If $\alpha > 1$, then $|\eta/\eta_\mathrm{co}| > 1$.
The electric field in the gap continues to grow, monotonically accelerating the particles until pair discharges ignite and short-circuit the electric field.
This is the main regime we study.

iii) \textbf{Anti-corotating flow.} For $\alpha < 0$, the corresponding charge density has the wrong sign to counter the growth of the electric field, resulting in similar growth and discharge dynamics as in the $\alpha > 1$ regime.

\subsection{Atmospheric Boundary}

The star supplies plasma for the circuit.
The atmosphere is an ample reservoir of electrons and protons, easily extracted by the unscreened parallel electric field.
The atmospheric charges screen the electric field close to the star so that $E_\parallel( h < 0 ) \approx 0$.
Only the electrons participate in the QED reactions;
positive current ($\alpha > 0$) lifts them into the magnetosphere.

The atmospheric plasma forms a thin, stratified layer on top of the star's surface with a scale height $H_p = p/g \rho$, where $p=p(\rho)$ is the pressure at a density $\rho$ and $g$ is the surface gravity:
$H_p \approx 6 \times 10^{-2} \cm$ for a fully ionized hydrogen layer with $\rho = 10^{-2} \,\mathrm{g}\,\mathrm{cm}^{-3}$ and $T \approx 10^5\,\mathrm{K}$, corresponding to the photosphere \citep{nattila2024b}.
The density drops to $n_\mathrm{co}$ in about 25 e-foldings, corresponding to a height of $25 H_p \sim 1.5 \cm$, about half the skin depth $c/\omega_{\mathrm{p,co}}$.

Gaps with unlimited atmospheric charge have been studied under the Space-Charge-Limited Flow scenario (SCLF;
\citealt{arons1979}).
The opposite case with no charge supply \citep{ruderman1975} is relevant, e.g., to magnetospheres above solid surfaces or around black holes.
These historical models have looked for steady-state gap solutions whereas we study time-dependent and intermittent gaps.
While there is conceptual difference between these models, we find it irrelevant for the gap dynamics, except that without charge supply pair cascades are ignited even when $0 < \alpha < 1$ \citep{timokhin2010, timokhin2013}.

\section{QED Processes}\label{sect:3}

QED processes enable photons and pair plasma to interact.
For moderate to large $b$, QED becomes nonlinear and the interaction cross sections are modified \citep[e.g.,][]{kiuru2026a,kiuru2026b}.
The quantum parameter $\chi$ determines the nonlinearity of the QED processes:%
\footnote{ Our $\chi$ is the magnetic analog of the nonlinearity parameter $\tilde{\chi} \equiv | p_\mu F^{\mu \nu} | / E_Q$ (where $F^{\mu\nu}$ is the electromagnetic tensor and $E_Q$ the critical electric field) used in laser plasma physics when a particle interacts with the electromagnetic plane-wave pulse \citep[e.g.,][]{fedotov2023}. The definitions agree up to $\chi^2 - \tilde{\chi}^2 = p^2 [b^2 - (E/E_Q)^2]$; they are exactly equal for massless particles ($p^2=0$) and when the first invariant vanishes, $F_{\mu \nu} F^{\mu \nu} = 0$. }
\begin{equation}
  \chi = \frac{| p_\mu G^{\mu \nu}|}{B_Q} \, ,
\end{equation}
where $p^\mu$ is the (dimensionless) four-momentum of the particle ($p^\mu = (\gamma, \gamma\vec{\beta})$ for leptons and $(x,x \nvec{\omega})$ for photons;
$x$ is the dimensionless photon energy in units of $m_e c^2$), and $G^{\mu\nu}$ is the Hodge dual of the electromagnetic tensor.
For leptons, it becomes
\begin{equation}\label{eq:chipm}
  \chi_\pm = \frac{\gamma}{B_Q} \sqrt{ (\vec{B}_\mathrm{eff}^{\pm})^2 - (\vec{\beta} \cdot \vec{B})^2 }
           \approx \gamma b \sin\psi \, ,
\end{equation}
where $\vec{B}_\mathrm{eff}^{\pm} \equiv \vec{B} - \vec{\beta} \times \vec{E}$ is the effective magnetic field and $\psi$ is the pitch angle w.r.t.\ the local $\vec{B}$ field.
The approximation in Eq.~\eqref{eq:chipm} corresponds to the ultra-relativistic limit $\gamma \gg 1$.
For photons, we similarly have
\begin{equation}
  \chi_x = \frac{x}{B_Q} \sqrt{ (\vec{B}_\mathrm{eff}^{x})^2 - (\nvec{\omega} \cdot \vec{B})^2  }
         \approx x b \sin\psi \, ,
\end{equation}
where $\vec{B}_\mathrm{eff}^{x} \equiv \vec{B} - \nvec{\omega} \times \vec{E}$ is now evaluated using the photon propagation direction $\nvec{\omega}$.

Prominent first-order QED processes include synchrotron emission (also known as nonlinear Compton;
$e^\pm \rightarrow e^\pm + \gamma$) and one-photon pair creation (nonlinear Breit--Wheeler;
$\gamma\rightarrow e^+ + e^-$).
The process rates (number of events per volume per time) scale as
\begin{equation}
  \dot{n}_{1 \rightarrow 2} \equiv \frac{\ud n}{\ud t}
                            \propto \frac{\alpha_\mathrm{em} b \sin\psi }{t_\mathrm{C}} n_1 T(\chi)
\end{equation}
for an interaction with one incoming and two outgoing particles.
Here $\alpha_\mathrm{em} \equiv e^2/(\hbar c) \approx 1/137$ is the fine-structure constant, $t_\mathrm{C} \equiv\lamC/c$ is the characteristic QED timescale, $\lamC \equiv \hbar/(m_e c)$ is the (reduced) electron Compton wavelength, $T(\chi)$ is a process-specific function (below, $T_\mathrm{syn}$ or $T_\mathrm{BW}$), and $n_1$ is the number density of the incoming particles.

\subsection{Photon Emission}

The energy-differential production rate of the synchrotron photons from an $e^\pm$-particle with a Lorentz factor $\gamma$ is \citep[e.g.,][]{kirk2009, berestetskii2012}
\begin{equation}
  \frac{\ud^2 N_{x} }{\ud t \, \ud x} =  \frac{2}{3}\frac{\alpha_\mathrm{em}}{t_\mathrm{C}} \chi_\pm^2 \frac{S(\gamma, x)}{x} \, ,
 \end{equation}
where $S(\gamma, x)$ is the spectral shape function.
The corresponding emission power is
\begin{equation}
  P_{\mathrm{syn}} \equiv m_e c^2 \int_0^\infty x \, \frac{\ud^2 N_{x} }{\ud t \, \ud x}  \, \ud x
                   = \frac{2}{3} \frac{\alpha_\mathrm{em}}{t_\mathrm{C}} \chi_\pm^2 g_\mathrm{Q}(\chi_\pm) m_e c^2 \, .
\end{equation}
A Gaunt factor $g_\mathrm{Q}(\chi_\pm) \equiv \int S \,\ud x \leq 1$ (with $S$ normalized so that $g_\mathrm{Q} \rightarrow 1$ for $\chi_\pm \ll 1$) suppresses this power;
we set $g_\mathrm{Q}= 1$, for simplicity, in our analytical estimates.

We obtain the differential optical depth (emission rate) for a single particle to emit a synchrotron photon, using the relation $\chi_\pm/\chi_x = \gamma/x$, as
\begin{equation}\label{eq:tau_syn}
  \frac{\ud \tau_\mathrm{syn}}{\ud t} \equiv \int_0^{\chi_\pm} \frac{\ud^2 N_{x}}{\ud t \, \ud \chi_x} \ud \chi_x
                                      = \frac{\alpha_\mathrm{em} b \sin\psi}{t_\mathrm{C}} T_\mathrm{syn}(\chi_\pm) \, .
\end{equation}
For $\chi_\pm \ll 1$, $T_\mathrm{syn}(\chi_\pm) \rightarrow 5/(2\sqrt{3}) \approx 1.44$.
In general, we can approximate $T_\mathrm{syn}$ (up to a relative error of $\sim 1\%$ for $\chi_\pm \lesssim 30$) as
\begin{equation}
  T_\mathrm{syn}(\chi_\pm) \approx \frac{5}{2\sqrt{3}} \left( 1 + \frac{16\sqrt{3}}{5} \chi_\pm + 1.1 \chi_\pm^2 \right)^{-1/6} \, .
\end{equation}
The linear coefficient $16\sqrt{3}/5 \approx 5.5$ reproduces the exact first-order expansion, $T_\mathrm{syn} = \frac{5}{2\sqrt{3}} ( 1 - \frac{8\sqrt{3}}{15}\chi_\pm ) + \mathcal{O}(\chi_\pm^2)$, while the quadratic coefficient is fitted.

The synchrotron emission process reduces to curvature radiation by substituting $\chi_\pm = \gamma b \sin\psi$ into $P_\mathrm{syn}$ and then replacing the pitch angle $\sin\psi \rightarrow \gamma r_\mathrm{g}/R_\mathrm{curv}$ (where $r_\mathrm{g} = c/\omega_B$ is the non-relativistic gyroradius), giving
\begin{equation}\label{eq:P_curv}
  \frac{P_\mathrm{curv}}{m_e c^2} = \frac{2}{3} \frac{\alpha_\mathrm{em} c}{\lamC} b^2 \frac{r_\mathrm{g}^2}{R_\mathrm{curv}^2} \gamma^4 \, .
\end{equation}

\subsection{Pair Production}

The differential optical depth for a gamma ray to pair-create via the one-photon Breit--Wheeler process is
\begin{equation}\label{eq:tau_bw}
  \frac{\ud \tau_\mathrm{BW}}{\ud t} = \frac{\alpha_\mathrm{em} b \sin\psi}{t_\mathrm{C}} T_\mathrm{BW}(\chi_x) \, ,
\end{equation}
where $T_\mathrm{BW}(\chi_x) \rightarrow 0.23\exp(-\frac{8}{3 \chi_x})$ for $\chi_x \ll 1$ \citep{erber1966} and $T_\mathrm{BW}(\chi_x) \rightarrow 0.38 \chi_x^{-1/3}$ for $\chi_x\gg 1$ \citep[e.g.,][]{kirk2009}.
We combine these limits and approximate $T_\mathrm{BW}$ (accurate to within $\sim 10\%$ for $\chi_x \lesssim 100$) as
\begin{equation}
  T_\mathrm{BW}(\chi_x) \approx 0.23 \exp\left(-\frac{8}{3 \chi_x} \right) \left( 1 + 0.22 \chi_x \right)^{-1/3} \, ,
\end{equation}
where the coefficient $0.22 \approx (0.23/0.38)^3$ ensures that both asymptotic limits are recovered.

The corresponding mean free path $l_\mathrm{mfp}$ (the average distance a photon travels before pair creation) is $l_\mathrm{mfp} \approx R_{\mathrm{curv}}\sin\psi_{\mathrm{a}}$, where $\psi_{\mathrm{a}}$ is the angle between the photon momentum and the local magnetic field after the photon has traveled the distance $l_\mathrm{mfp}$.
Since $\chi_{x} \approx x b \sin \psi$, we get $l_\mathrm{mfp} \approx R_{\mathrm{curv}} \chi_{x}^{\mathrm{a}}/xb$,
where $\chi_{x}^{\mathrm{a}}$ is defined such that $\tau_\mathrm{BW}(\chi_{x}^{\mathrm{a}}) = 1$. %
An approximate formula for estimating the photon's mean free path (in the $\chi_x \ll 1$ limit) is%
\footnote{%
Here we make use of an approximate formula for $\tau_\mathrm{BW}$ that can be obtained from Eq.~\eqref{eq:tau_bw} (see also Appendix B of \citealt{timokhin2019} with somewhat similar approximations):
\begin{align*}
\begin{split}
&\tau_\mathrm{BW}(\chi_{x}^{\mathrm{a}}) = \int_{0}^{l_{\mathrm{mfp}}}\frac{\alpha_\mathrm{em}}{t_\mathrm{C}} \frac{\chi_x}{x} T_\mathrm{BW}(\chi_x) \frac{\ud l}{c}
  \approx 0.23 \frac{\alpha_\mathrm{em}}{\lamC} \int_{0}^{l_{\mathrm{mfp}}} \frac{\chi_{x}}{x} e^{-\frac{8}{3\chi_{x}}} \ud l \\
&\approx 0.23 \frac{\alpha_\mathrm{em} R_{\mathrm{curv}}}{\lamC x^{2} b} \int_{0}^{\chi_{x}^{\mathrm{a}}} \chi_{x} e^{-\frac{8}{3\chi_{x}}} \ud \chi_{x}
= 0.23 \frac{\alpha_\mathrm{em} R_{\mathrm{curv}}}{\lamC x^{2} b} (\chi_{x}^{\mathrm{a}})^{2} E_{3}\Big( \frac{8}{3 \chi_{x}^{\mathrm{a}}}\Big) \\
&\approx 0.086 \frac{\alpha_\mathrm{em} R_{\mathrm{curv}}}{\lamC x^{2} b} (\chi_{x}^{\mathrm{a}})^{3} e^{-\frac{8}{3\chi_{x}^{\mathrm{a}}}} \, ,
\end{split}
\end{align*}
where $E_{3}$ is the exponential integral of order $3$ and for small argument $z$ we have $z^{2} E_{3}(\frac{8}{3z}) \approx \frac{3}{8} z^{3} e^{-8/(3z)} + \mathcal{O}(z^{4}e^{-8/(3z)})$.
}
\begin{align} \label{eq:lmfp_pair}
\begin{split}
l_\mathrm{mfp} &\approx \frac{8}{9} \frac{R_{\mathrm{curv}}}{xb}
W \Big[ \frac{2}{5} \Big(\frac{ \lamC x^{2}b  }{ \alpha_\mathrm{em} R_\mathrm{curv}} \Big)^{-1/3}\Big]^{-1}  \\
&\approx \frac{8}{9} \frac{R_{\mathrm{curv}}}{xb}  \bigg\{ \ln \left[ \left(\frac{\alpha_\mathrm{em} R_\mathrm{curv}}{16 \lamC x^{2}b} \right)^{1/3} \right] \\
& \quad\quad\quad\quad\quad -\ln\ln \left[ \left(\frac{\alpha_\mathrm{em} R_\mathrm{curv}}{16 \lamC x^{2}b} \right)^{1/3} \right]
\bigg\}^{-1} \, ,
\end{split}
\end{align}
where $W$ is the Lambert $W$ function;
for large argument $z$, $W(z) \approx \ln z - \ln\ln z$.
We have $l_\mathrm{mfp} \sim 10^{-2} R_\mathrm{pc}$ for the radio-pulsar parameters when $x \sim x_\mathrm{curv}$ (see Eq.~\ref{eq:xcurv}).

\subsection{Neglected Processes}\label{sect:higher_order}

We ignore other first-order processes besides the synchrotron and one-photon pair creation.
These include, e.g., synchrotron absorption ($e^\pm +\gamma\rightarrow e^\pm$) and one-photon pair annihilation ($e^- + e^+ \rightarrow \gamma$).
The synchrotron self-absorption rate is small because the high-energy leptons preferentially absorb low-energy photons---the number density of such target particles is small, resulting in low process rates.
The one-photon pair annihilation cross section, on the other hand, is $\sigma_\mathrm{ann} \approx 0$ for $b \lesssim 1$, rendering the process unlikely.

We also ignore all the higher-order processes.
These include, e.g., QED processes with two vertices and $2+2$ external legs, such as Compton scattering ($e^\pm + \gamma \rightarrow e^\pm + \gamma$), two-photon pair creation ($\gamma + \gamma \rightarrow e^- + e^+$), and pair annihilation ($e^- + e^+ \rightarrow \gamma + \gamma'$).
These processes are not thought to be dominant for radio pulsars.
However, they can be important for systems with $b \ll 1$, such as millisecond pulsars (Salmi \& N\"attil\"a, in prep.).
Other effects, such as photon polarization, if included, enhance the pair yield in the cascade by a few percent \citep{song2024}.

\section{Particle Dynamics in the Gap}\label{sect:4}

We consider individual particle dynamics in the gap.
The time evolution of the proper velocity (the spatial part of the four-velocity) $\vec{u} = \gamma \vec{\beta} c$ of a charged particle, such as an electron in the stellar atmosphere, follows
\begin{equation}
  \frac{\ud \vec{u} }{\ud t} = \frac{q}{m_e} \left( \vec{E} + \frac{\vec{v}}{c} \times \vec{B} \right) - \frac{\vec{F}_\mathrm{rad}}{m_e} \, ,
\end{equation}
where $q = -e$ for electrons and $q = +e$ for positrons (with $e > 0$ the elementary charge), and $\vec{F}_\mathrm{rad}$ is the radiative drag force.
In the strongly magnetized limit the dynamics reduce to a one-dimensional beads-on-a-wire approximation with a spatial coordinate $h$ along the magnetic field $\vec{B}$.
In addition, for relativistic particle (upward-)motion $u \equiv \beta \gamma c \rightarrow \gamma c$ and $P_\mathrm{rad} = \vec{F}_\mathrm{rad} \cdot \vec{\beta} c \rightarrow F_\mathrm{rad} c$.
Then, the particle dynamics equation reduces to
\begin{equation}
  \frac{\ud \gamma}{\ud t} = \frac{q E_\parallel}{m_e c} - \frac{P_\mathrm{rad}}{m_e c^2} \, .
\end{equation}

The parallel electric field at the polar cap can be approximated with a linear-voltage gap \citep[e.g.,][]{daugherty1982, timokhin2010}
\begin{equation}
E_\parallel =
\begin{cases}
    -\frac{\Delta V}{H_\mathrm{gap}}, \quad & h \leq H_\mathrm{gap} \\
    0,                  \quad  &  h > H_\mathrm{gap} \, ,
\end{cases}
\end{equation}
where $H_\mathrm{gap}$ is the extent of the region and $h = 0 $ at the star's surface.
$H_\mathrm{gap} = R_\mathrm{pc}$ is a typical choice.

Curvature radiation provides the radiative force with an emission power
\begin{equation}\label{eq:dgamma_curv}
  \frac{\ud \gamma}{\ud t} \Bigg|_\mathrm{curv} = -\frac{P_\mathrm{curv}}{m_e c^2} \, .
\end{equation}
The characteristic photon energy%
\footnote{ The characteristic energy of a synchrotron photon is $x_\mathrm{syn} = \frac{3}{2} b \gamma^2 \sin\psi$; the curvature energy is obtained by substituting $\sin\psi \rightarrow \gamma r_g/R_\mathrm{curv}$. }
is
\begin{equation}\label{eq:xcurv}
  x_\mathrm{curv} \equiv \frac{\hbar \omega_\mathrm{curv}}{m_e c^2}
                  = \frac{3}{2} b \frac{r_g}{R_\mathrm{curv}} \gamma^3 \, ,
\end{equation}
where $r_g \sim 10^{-9} \cm$, so that $r_g/R_\mathrm{curv} \sim 10^{-15}$.

The particle dynamics in terms of distance from the surface $h$ follow from $\ud h \approx c \ud t$:
\begin{equation}\label{eq:gam_forces}
  \frac{ \ud \gamma }{\ud h} = \frac{q E_\parallel}{m_e c^2} - \frac{F_\mathrm{rad}}{m_e c^2}
                             \approx - \frac{q \Delta V}{m_e c^2 H_\mathrm{gap}}  - \frac{2}{3} \frac{\alpha_\mathrm{em}}{\lamC} \frac{r_g^2}{R_{\mathrm{curv}}^2} b^2 \gamma^4 \, .
\end{equation}

The maximum Lorentz factor of an electron experiencing the full potential drop across the polar cap (i.e., $h$ goes from $0$ to $H_\mathrm{gap}$) is obtained when $F_\mathrm{rad} \rightarrow 0$,
\begin{equation}\label{eq:gam_gap}
  \gamma_{\mathrm{gap}} = \int_{0}^{H_\mathrm{gap}} \frac{\ud \gamma }{\ud h} \ud h
                        = \frac{e \Delta V}{m_e c^2}
                        \sim \frac{e B_\star \beta_\mathrm{rot} R_\mathrm{pc}}{2 m_e c^2} \, .
\end{equation}
We have $\gamma_\mathrm{gap} \sim 10^{7}$.

The steady-state Lorentz factor ($\ud \gamma /\ud t = 0$) follows from balancing acceleration and radiative drag.
The corresponding radiation-threshold Lorentz factor is then \citep{daugherty1982},
\begin{align}\begin{split}\label{eq:gam_rad}
  \gamma_{\mathrm{rad}}
  &= \left( \frac{3}{2} \frac{e \Delta V}{m_e c^2} \frac{B_Q^2}{B_\star^2} \frac{R_\mathrm{curv}^2}{r_g^2} \frac{\lamC}{\alpha_\mathrm{em} H_\mathrm{gap}} \right)^{1/4} \\
 & = \gamma_\mathrm{gap}^{1/4} \, b^{-1/2} \left(\frac{3 \lamC}{2 \alpha_\mathrm{em} H_\mathrm{gap}} \right)^{1/4} \left( \frac{r_g}{R_\mathrm{curv}} \right)^{-1/2}
\end{split}\end{align}
and it is obtained over a distance%
\footnote{%
The differential equation for the Lorentz factor $\gamma$ and distance $l$ is of the form $\ud \gamma/\ud l = \mathcal{A} - \mathcal{B} \gamma^4$,
where the auxiliary constants are $\mathcal{A} \equiv \gamma_\mathrm{gap}/H_\mathrm{gap}$ and $\mathcal{B} \equiv (2/3)(\alpha_\mathrm{em}/\lamC)(b^2 r_g^2 / R_\mathrm{curv}^2)$.
Then, $\ud l = \ud \gamma/(\mathcal{A} - \mathcal{B}\gamma^4)$,
and so
\begin{equation}
  l = \int \ud l = \int \frac{\ud \gamma}{\mathcal{A} - \mathcal{B}\gamma^4} \approx \frac{\gamma}{\mathcal{A}} + \frac{\mathcal{B} \gamma^5}{5 \mathcal{A}^2} + \mathcal{O}(\gamma^9) \, ,
\end{equation}
solved via Taylor expansion of the integral (convergent for $\gamma < \gamma_\mathrm{rad}$, which we use as an order-of-magnitude estimate near the limit).
Physically, $\mathcal{A} \gg \mathcal{B}\gamma^4$ holds throughout most of the acceleration phase, i.e., the radiative losses remain subdominant until the particle approaches the radiation-reaction limit.
} of about $(\gamma_\mathrm{rad}/\gamma_\mathrm{gap}) H_\mathrm{gap}$.
For our parameters, $\gamma_\mathrm{rad} \sim 10^7$ and $\gamma_\mathrm{rad}/\gamma_\mathrm{gap} \approx 0.6$.

The accelerated pairs emit curvature photons with a characteristic energy $x_\mathrm{curv}$.
These photons are prone to pair creation when their energy exceeds the threshold, $x \gtrsim x_{\mathrm{thr}} \simeq 2$;
this happens when the particle's energy exceeds
\begin{equation}
\gamma_\mathrm{thr} \sim \left( \frac{4}{3} \frac{ R_\mathrm{curv} }{ b r_g}\right)^{1/3} \, ,
\end{equation}
which is $\gamma_\mathrm{thr} \sim 3 \times 10^5$.
Copious pair production occurs when $\gamma_\mathrm{rad} \gg \gamma_\mathrm{thr}$.

Since $\chi_x$ grows with the propagation angle as $\sin\psi \approx l/R_\mathrm{curv}$, a photon converts only if its mean free path fits inside the gap, $l_\mathrm{mfp}(x) \lesssim H_\mathrm{gap}$.
With $l_\mathrm{mfp} \approx R_\mathrm{curv} \chi_{x}^{\mathrm{a}}/(x b)$, this defines the minimum energy of a converting photon,
$ x_\mathrm{min} \approx \chi_{x}^{\mathrm{a}} R_\mathrm{curv} /(H_\mathrm{gap} b) \approx 500$,
for the fiducial parameters and ignoring the weak dependency of $\chi_x^{\mathrm{a}}$ on $x$.
The corresponding minimum injection energy for the resultant pairs is, therefore,
\begin{equation}\label{eq:gam_inj}
  \gamma_\mathrm{min} = \frac{x_\mathrm{min}}{2} \approx 250 \, .
\end{equation}

We conclude that there is a hierarchy of energy scales
\begin{equation}
1 \ll \gamma_{\mathrm{min}} \ll  \gamma_\mathrm{thr} < \gamma_\mathrm{rad} \lesssim \gamma_\mathrm{gap} \, ,
\end{equation}
which also needs to be fulfilled in numerical simulations.

\section{Discharge Dynamics}\label{sect:discharge}

Particles energized in the gap regions can trigger explosive plasma production and intermittent limit-cycle behavior.
We construct a concurrency model for the discharge dynamics and plasma generation in the gap regions.
See also Appendix~\ref{app:discharge} for the full derivation of the analytic expressions.

\subsection{Characteristic Timescales}

The gap height $H_\mathrm{gap} \sim R_\mathrm{pc}$ determines the slowest of the gap timescales, the light-crossing time
\begin{equation}
t_\mathrm{esc} \equiv \frac{H_\mathrm{gap}}{c} \sim 5 \times 10^{-7} \sec \, .
\end{equation}

Particle acceleration sets the fastest discharge timescale.
The energization is rapid,
\begin{equation}
\frac{\mathrm{d} \gamma}{\mathrm{d} t} \sim \frac{\Delta \gamma}{t_\mathrm{acc}} \approx \frac{q E_\parallel}{m_e c} \, ,
\end{equation}
with a timescale
\begin{equation}
t_\mathrm{acc} \sim \left| \frac{\Delta \gamma m_e c}{e E_\parallel} \right| \approx \frac{m_e c}{e \beta_\mathrm{rot} B_\star} = \beta_\mathrm{rot}^{-1}\omega_B^{-1} \sim 10^{-14} \sec \, ,
\end{equation}
where $E_\parallel \approx -\beta_\mathrm{rot} B_\star$ and $\Delta \gamma = 1$ (i.e., energy gain of $m_e c^2$).

Curvature-photon emission and one-photon pair creation proceed on intermediate timescales.
High-energy curvature photons (with $x \gtrsim 2$, generated by pairs with $\gamma > \gamma_\mathrm{thr}$) travel, on average, for a duration of $t_\pm = l_\mathrm{mfp}/c \sim 10^{-8} \sec$ (where $l_\mathrm{mfp}$ is given by Eq.~\eqref{eq:lmfp_pair}, evaluated at $x = x_\mathrm{curv}(\gamma_\mathrm{rad})$) before becoming an $e^\pm$-pair.
The characteristic curvature emission time can be estimated from Eq.~\eqref{eq:tau_syn} as%
\footnote{
Equivalently, the timescale can be estimated (up to an order-unity factor), by assuming an emission power $P_\mathrm{curv}$ and a characteristic photon energy of $x_\mathrm{curv}$, as
$t_x \sim x_\mathrm{curv} m_e c^2 / P_\mathrm{curv}$.
}
\begin{equation}
\frac{\ud \tau_\mathrm{curv}}{\ud t} \sim t_x^{-1} \approx \alpha_\mathrm{em} b\frac{c}{\lamC} \frac{r_g}{R_\mathrm{curv}} \gamma_\mathrm{rad}
\end{equation}
to give
\begin{equation}
t_x \sim \frac{\lamC}{\alpha_\mathrm{em} b c} \frac{R_\mathrm{curv}}{r_g} \frac{1}{\gamma_\mathrm{rad}} \sim 5 \times 10^{-10} \sec \, .
\end{equation}
Hence, for radio pulsars, $t_\mathrm{acc} \ll t_x \lesssim t_\pm \ll t_\mathrm{esc}$.
In the discharge model below we idealize $t_x \sim t_\pm$.

The gap with $|E_\parallel/E_\mathrm{rot}| > 0$ remains exposed until a current of $j_\pm = e n_\pm c$ screens it.
The screening time can be estimated from Maxwell's equation
\begin{equation}
\partial_t E_\parallel \sim \frac{E_\parallel}{t_\mathrm{screen}} \approx - 4\pi j_\pm \, ,
\end{equation}
where we have ignored $j_\mathrm{m}$, since $j_\mathrm{m} \ll j_\pm$ during the active screening phase. Then,
\begin{equation}
t_\mathrm{screen} \sim \left| \frac{E_\parallel}{4\pi j_\pm} \right| \approx  \frac{t_\mathrm{esc}}{M_\pm'} \, ,
\end{equation}
where $M_\pm' = n_\pm/n_\mathrm{co} \gg 1$ is the plasma multiplicity at the onset of screening (defined more rigorously in the next section).

In addition to the discharge timescales, the plasma has two kinetic timescales.
The gyro timescale satisfies $\omega_B^{-1} \ll t_\mathrm{acc}$, which renders the particle motion 1D along the $\vec{B}$ field.
The nominal plasma timescale satisfies $\omega_{\mathrm{p,co}}^{-1} \approx t_\mathrm{screen}$, indicating that gap screening is inherently nonlinear:
the plasma responds on the same timescale as the field evolution.

The discharge dynamics then has a hierarchy of timescales set as
\begin{equation}\label{eq:timescales}
\omega_B^{-1}
\ll
t_\mathrm{acc}
\ll
t_\mathrm{screen}
\sim
\omega_{\mathrm{p,co}}^{-1}
\lesssim
t_x
\lesssim t_\pm
\ll
t_\mathrm{esc} \, .
\end{equation}

\subsection{Discharge Dynamics Model}

We express the pair number density $n_\pm$ (of all charged particles) and photon number density $n_x$ (of photons above the pair-creation threshold) inside the discharge region as
\begin{align}
\frac{\mathrm{d} n_\pm}{\mathrm{d} t} &= 2\frac{n_x}{t_\pm}  -\frac{n_\pm}{t_\mathrm{esc}}  \\
\frac{\mathrm{d} n_x}{\mathrm{d} t}   &= -\frac{n_x}{t_\pm} + \frac{n_\pm}{t_x}\mathcal{H}\left( \left| \frac{ E_\parallel}{E_\mathrm{rot}}\right| - 1\right)  -\frac{n_x}{t_\mathrm{esc}} \, ,
\end{align}
where $\mathcal{H}$ is the Heaviside step function; it models the threshold electric field needed for pair creation, with the limit placed, for simplicity, at the full rotational field $|E_\parallel/E_\mathrm{rot}| = 1$.

The electric field in the gap evolves according to Maxwell's law
\begin{equation}
\frac{\mathrm{d} E_\parallel}{\mathrm{d} t} \approx 4\pi (j_\mathrm{m} - j_\pm) \, .
\end{equation}
The current from the pair plasma, $j_\pm = \sum_{s} q_{s} n_s \beta_{\parallel,s} c$, acts to screen the exposed electric field.
Assuming the two species flow in opposite directions with the same speed $|\beta_{\parallel}|$, we have $|j_\pm| = e n_{\pm}|\beta_{\parallel}|c$.

The current dynamics follow from the 1D Vlasov equation for the particle distribution $f_s(h,u,t)$ along the magnetic field:
\begin{equation}
\frac{\partial f_s}{\partial t} = - c\beta \frac{\partial f_s}{\partial h} - \frac{q_s E_\parallel}{m_e} \frac{\partial f_s}{\partial u} \, .
\end{equation}
Here we have ignored the source term from the particle generation.
The current equation follows from differentiating its definition and substituting the Vlasov equation \citep[see also][]{tolman2022},
\begin{align}
\begin{split}
\frac{\partial j_\pm}{\partial t} &= \sum_{s} q_s  \int c\beta \frac{\partial f_s}{\partial t}  \mathrm{d} u \\
                                  &= \sum_{s}\left( -q_s \int c^2\beta^2 \frac{\partial f_s}{\partial h}  \mathrm{d}u + \frac{q^2_s}{m_e} \int c E_\parallel f_s \frac{\partial \beta}{\partial u}  \mathrm{d} u \right)\\
                                  &\approx  -\frac{c}{L}j_\pm + \frac{e^2 E_\parallel n_\pm}{m_{e}} \left\langle\frac{1}{\gamma^3} \right\rangle \, ,
\end{split}
\end{align}
where
$n_\pm = \int f \, \mathrm{d}u$,
$\left\langle \gamma^{-3} \right\rangle \equiv n_\pm^{-1} \int f \gamma^{-3} \mathrm{d}u$, $f = \sum_s f_s$ and
we have approximated $c \partial_h \rightarrow c/L$ (where $L \sim H_\mathrm{gap}$ is the characteristic length scale of variations in $j_\pm$), and $\beta \rightarrow 1$.

Next, we express the number densities as $n_\pm \rightarrow m_\pm n_\mathrm{co}$ and $n_x \rightarrow m_x n_\mathrm{co}$ via the multiplicities $m_\pm$ and $m_x$.
Then, the current can be expressed as $j_\pm \rightarrow j e n_\mathrm{co} c$, where $|j| \equiv m_\pm |\beta_{\parallel}|$ and $j_\mathrm{m} \rightarrow \alpha j_\mathrm{co}$.
Similarly, we express the electric field as a function of the rotational field, $E_\parallel \rightarrow \varepsilon |E_\mathrm{rot}|$, where $\varepsilon$ is the dimensionless electric field.
The system of coupled ODEs modeling the discharge region is:
\begin{align}
\dot{m}_\pm &= 2 \frac{m_x}{t_\pm}  -\frac{m_\pm}{t_\mathrm{esc}} \label{eq:dis1} \\
\dot{m}_x   &= -\frac{m_x}{t_\pm} + \frac{m_\pm}{t_x}\mathcal{H}(|\varepsilon| - 1)  -\frac{m_x}{t_\mathrm{esc}} \\
\dot{\varepsilon}   &= \frac{\alpha - j}{t_\mathrm{esc}} \\
\dot{j} &= -\frac{j}{t_\mathrm{esc}} + \frac{m_\pm \varepsilon}{t_\mathrm{acc}} \left\langle\frac{1}{\gamma^3} \right\rangle \, , \label{eq:dis4}
\end{align}
where overdots denote time derivatives ($\dot{Q} \equiv \ud Q/\ud t$).

The system also requires an expression for the evolution of the plasma-inertia measure $\langle \gamma^{-3} \rangle$, as it changes during the discharge cycle.
We have tested solving such an extended model, obtainable via a moment expansion of the Vlasov equation, but found the solutions qualitatively similar to the simplified case with $\partial_t \langle \gamma^{-3} \rangle = 0$.
Here we choose to approximate the inertia measure as a constant fixed to the ignition-phase value $\langle \gamma^{-3} \rangle_\mathrm{ign} \sim \gamma_\mathrm{min}^{-3}$, consistent with the value measured in the numerical simulations, while noting the full variation observed in kinetic simulations (see Sect.~\ref{sect:sims}).

The coupled system has five qualitatively different phases that we discuss next; the last three overlap in time.

\subsection{Limit Cycle Phases}\label{sect:phases}

\textbf{Loading.}
We analyze the early loading phase of the discharge cycle with $t \lesssim \sqrt{t_\pm t_x}$.
The system starts from a state where the electric field has reached a critical value $\varepsilon \approx 1$, the plasma density is low ($m_\pm(t = 0) \approx  1$), and no high-energy photons exist in the gap region ($m_x(t = 0) = 0$).
Then, the photon number density $m_x$ evolves according to
\begin{equation}
\dot{m}_x  = \frac{m_\pm}{t_x} \, ,
\end{equation}
with a linear-growth solution of
\begin{equation}
m_x(t) =  \frac{t}{t_x} \, .
\end{equation}
During this time, the pair density increases only slightly according to
\begin{equation}
\dot{m}_\pm  = 2\frac{m_x(t)}{t_\pm} = \frac{2t}{t_x t_\pm} \, ,
\end{equation}
with a solution
\begin{equation}
    m_\pm(t)  =  1 + \frac{t^2}{t_\pm t_x} \, .
\end{equation}
The system remains in the loading phase until $m_x(t) \sim 1$, which is reached in $t \sim t_x$.
The phase does not exist at all for a pre-populated gap.

\textbf{Ignition.}
The ignition phase starts when the photon multiplicity has become comparable to the plasma multiplicity, $m_x(t) \sim 1$.
The phase lasts until the gap is screened (i.e., while $|\varepsilon| \geq 1$).
The dynamical equations for $t \ll t_\mathrm{esc}$ are
\begin{align}
\dot{m}_\pm &\approx 2 \frac{m_x}{t_\pm}  \, , \\
\dot{m}_x   &\approx -\frac{m_x}{t_\pm} + \frac{m_\pm}{t_x} \, .
\end{align}
Importantly, for $t \gtrsim t_\pm$, both solutions simplify to
\begin{equation}
m_{\pm/x}(t)  \propto \exp\left( \frac{t}{t_\pm} \right) \, ,
\end{equation}
when $t_\pm \sim t_x$.
During this phase, the multiplicities experience exponential growth in the exposed gap.
The phase stops abruptly when the pair multiplicity becomes high, $m_\pm \gg 1$, and the resulting strong current $j_\pm \propto m_\pm$ shorts the exposed electric field, $\varepsilon = 1 \rightarrow 0$.

\textbf{Oscillations.}
The sudden screening of the gap leads to electric field oscillations because the plasma-induced current has inertia and does not instantly respond to the electric field changes.
For $m_\pm$ and $m_x$ these fluctuations are negligible because they average out and decouple from the dynamical equations.
However, the oscillations are, in general, important because they excite plasma waves in the gap region, which can escape and become the observed radio emission.
The oscillations are analyzed also in \citet{levinson2005, tolman2022, cruz2021b, okawa2024}.

Combining the equations for $\dot{\varepsilon}$ and $\dot{j}$ and assuming a constant plasma multiplicity $m_\pm(t) \sim m_\pm$ yields a damped harmonic oscillator with constant forcing
\begin{equation}
\ddot{\varepsilon} + \frac{ \dot{\varepsilon} }{t_\mathrm{esc}} +  \omega_\mathrm{osc}^2 \varepsilon = \frac{\alpha}{t_\mathrm{esc}^{2}} \, ,
\end{equation}
where the oscillation frequency
\begin{equation}\label{eq:omega_osc}
\omega_\mathrm{osc}^2 = \frac{m_\pm}{t_\mathrm{acc} t_\mathrm{esc}} \left\langle\frac{1}{\gamma^3} \right\rangle
= \omega_{\mathrm{p},\mathrm{co}}^2 m_\pm \langle \gamma^{-3} \rangle
\, .
\end{equation}
The discharge rings at the plasma frequency of the gap, suppressed by the effective inertia $\gamma^3 m_e$ of the pairs.
The oscillator is underdamped with a damping ratio $\zeta = (2\omega_\mathrm{osc} t_\mathrm{esc})^{-1} \ll 1$ for any realistic $m_\pm$.
The amplitude damping time of the oscillations is $t_\mathrm{damp} = 2 t_\mathrm{esc}$.
Therefore, a broad-band burst of radio emission at a characteristic angular frequency of $\sim \omega_\mathrm{osc}$ and duration $\sim t_\mathrm{esc}$ is expected from the discharges with any realistic multiplicity.

During the oscillation phase, $\varepsilon$ oscillates on the fast timescale $\omega_\mathrm{osc}^{-1} \ll t_\mathrm{esc}$, so the current responds only to the cycle-averaged field and relaxes to the quasi-static solution
\begin{equation}\label{eq:jpm}
j(t) \approx m_\pm \varepsilon \frac{t_\mathrm{esc}}{t_\mathrm{acc}} \left\langle\frac{1}{\gamma^3} \right\rangle \, .
\end{equation}
Therefore, we can compare this expression to the earlier definition of $|j| = m_\pm |\beta_\parallel|$ and identify the bulk velocity as
\begin{equation}\label{eq:beta-parallel}
|\beta_\parallel| \approx |\varepsilon| \frac{t_\mathrm{esc}}{t_\mathrm{acc}} \left\langle\frac{1}{\gamma^3} \right\rangle \, .
\end{equation}
A good proxy for the current is then $j \approx C_1 \varepsilon m_\pm$, where $C_1$ is a constant.
The linear-response estimate is valid only while $|\beta_\parallel| < 1$;
the current saturates at $|j| \rightarrow m_\pm$ if the bulk flow becomes relativistic.

\textbf{Saturation.}
The sudden screening of the gap leads to saturation of the plasma and photon generation.
The screened electric field halts curvature-photon production since $\mathcal{H}(|\varepsilon| - 1) = 0$.
The system simplifies to
\begin{align}
\dot{m}_\pm &= 2\frac{m_x}{t_\pm}  -\frac{m_\pm}{t_\mathrm{esc}} \, , \\
\dot{m}_x   &= -\frac{m_x}{t_\pm} -\frac{m_x}{t_\mathrm{esc}} \, ,
\end{align}
with an exponential-decay solution for the photons,
\begin{equation}\label{eq:mx_sat}
m_x(t) = M_x' \exp\left( -\frac{t}{t_\mathrm{esc}} - \frac{t}{t_\pm}\right) \, ,
\end{equation}
and a round-topped solution for the pairs,
\begin{equation}\label{eq:mp_sat}
m_\pm(t) = m_x(t)
\left\{
\exp\left(\frac{t}{t_\pm} \right) \left[ \frac{ M_\pm' }{ M_x' } + 2 \right] - 2
\right\}
\, ,
\end{equation}
where $M_x'$ and $M_\pm'$ are the photon and pair multiplicities at the onset of screening, respectively.
The peak of $m_\pm(t)$ occurs after a delay of $t_\mathrm{d} = t_\pm \ln \left\{ 2 M_x' (t_\mathrm{esc} + t_\pm)/[ t_\pm (M_\pm' + 2 M_x')] \right\}$ from the time of the screening, provided $2 M_x'/M_\pm' > t_\pm/t_\mathrm{esc}$; otherwise the maximum occurs at the onset of screening.

As a caveat, this phase models the pair cascade only up to the second generation.
In contrast, our PIC simulations find that the cascade generates up to six generations of pairs (cf.\ the five to eight generations reported by \citealt{timokhin2019}).
Therefore, a realistic saturation phase can last longer and produce more pairs.

\textbf{Decay.}
In the last decay phase, pair production fades as most high-energy photons are already converted to pairs.
At late times ($t \gg t_\pm$), the pairs decay as
\begin{equation}\label{eq:m_ph5}
m_\pm(t) \approx (M_\pm' + 2 M_x')  \exp\left( -\frac{t}{t_\mathrm{esc}} \right ) \approx M_\pm \exp\left( -\frac{t}{t_\mathrm{esc}} \right ) \, ,
\end{equation}
where we have introduced the peak multiplicity $M_\pm \approx M_\pm' + 2 M_x'$ reached during the discharge.

As the particle density fades away, the gap starts to reform and the electric field grows as $\dot{\varepsilon} \approx \alpha/t_\mathrm{esc}$ with a solution of $\varepsilon(t) = \alpha (t/t_\mathrm{esc})$.
The gap restoration starts when $|j|  \approx m_\pm |\beta_\parallel| < |\alpha|$; because the remaining pairs are relativistic at this stage ($|\beta_\parallel| \rightarrow 1$), this gives $m_\pm \lesssim |\alpha| \sim$ a few as the threshold below which the gap starts to re-grow.

\subsection{Macroscopic Properties}

The peak multiplicity $M_\pm$ of the first- and second-generation pairs is set by the screening of the accelerating electric field during ignition (see Appendix.~\ref{app:discharge}),
\begin{equation}\label{eq:Mpm}
  M_{\pm} \sim 10 \frac{t_\mathrm{acc}}{t_x\, \gamma_\mathrm{min}^{-3} } \, .
\end{equation}
For the typical pulsar parameters, we obtain $M_\pm \sim 5 \times 10^3$.

The total cycle duration $t_\mathrm{cycle}$---obtained from the condition $m_\pm(t_\mathrm{cycle}) = |\alpha|$ in Eq.~\eqref{eq:m_ph5}---is dominated by the decay phase and is given as
\begin{equation}\label{eq:t_cycle}
  t_\mathrm{cycle} \approx \ln \left( \frac{M_\pm}{|\alpha|} \right) \, t_\mathrm{esc} \, .
\end{equation}
The time-averaged multiplicity is
\begin{align}
\begin{split}
\langle m_\pm \rangle &= \frac{1}{t_\mathrm{cycle}}  \int_0^{t_\mathrm{cycle}} m_\pm(t) \,\mathrm{d}t
\approx \frac{M_\pm}{\ln M_\pm}\, ,
\end{split}
\end{align}
where we have used Eq.~\eqref{eq:m_ph5} for $m_\pm(t)$.

\subsection{Numerical Solution}\label{sect:ode_solution}

\begin{figure}[t!]
\centering
\includegraphics[clip, trim=0.0cm 0.0cm 0.0cm 0.0cm, width=8.5cm]{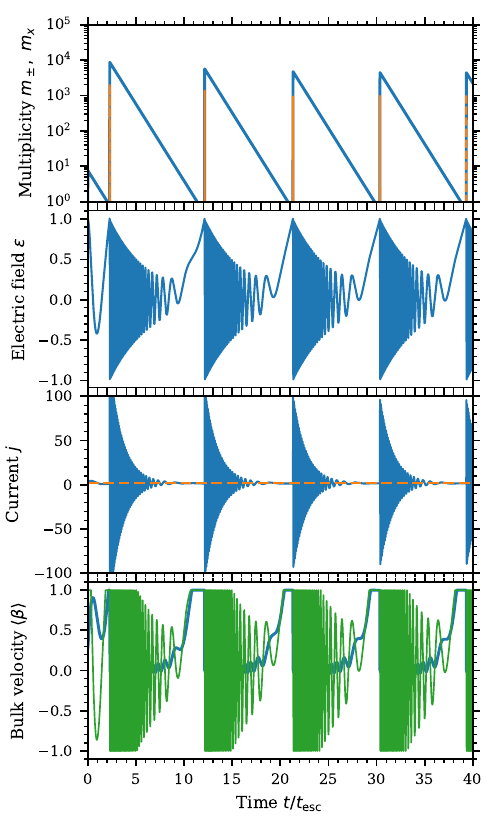}
\caption{\label{fig:discharge_model}
Numerical solution of the analytic discharge model for the fiducial parameters $\alpha = 2$, $t_x = t_\pm = 10^{-3} t_\mathrm{esc}$, $t_\mathrm{acc} = 4\times 10^{-8} t_\mathrm{esc}$, and $\langle \gamma^{-3} \rangle = 8 \times 10^{-8}$.
Top panel shows the plasma (blue curve) and photon (orange dashed) multiplicities.
Second panel shows the electric field.
Third panel shows the electric current (blue) and $\alpha$ (orange dashed).
Bottom panel shows the bulk flow velocity $\langle \beta \rangle = \pm j/m_\pm$ (blue) and its proxy $\varepsilon  (t_\mathrm{esc}/t_\mathrm{acc}) \langle \gamma^{-3} \rangle$ (green); both are capped at the physical bound $|\beta| = 1$.
}
\end{figure}

The full numerical solution of the ODE system is shown in Fig.~\ref{fig:discharge_model}.
We use the Radau method (implicit Runge--Kutta method of the Radau IIA family of order 5) to solve the stiff ODE system.
The solution is numerically straightforward to obtain and is not sensitive to the initial conditions.

The numerical result exhibits all the phases described in the Sect.~\ref{sect:phases} and has limit-cycle behavior.
The measured peak multiplicity, $M_\pm \approx 5 \times 10^3$ (with a cycle-to-cycle scatter of $\sim\!20\%$), matches the estimate of Eq.~\eqref{eq:Mpm}.
The measured cycle length $(9.1 \pm 0.1)\, t_\mathrm{esc}$ is comparable to the estimated $t_\mathrm{cycle} \approx \ln(M_\pm / |\alpha|)\, t_\mathrm{esc} \approx 7.8\, t_\mathrm{esc}$.
We have also verified that replacing the $t_\pm = t_x$ idealization with the physical value $t_\pm \approx 2 \times 10^{-2}\, t_\mathrm{esc}$ (and retaining $t_x = 10^{-3} t_\mathrm{esc}$) leaves the peak multiplicity essentially unchanged:
the slower cascade growth rate is compensated by the larger photon-to-pair ratio at the moment of screening, which boosts the post-screening pair yield.
The simplification is therefore robust.

\begin{figure*}[tp!]
\centering
\includegraphics[clip, trim=0.0cm 0.0cm 0.1cm 0.0cm, height=21cm]{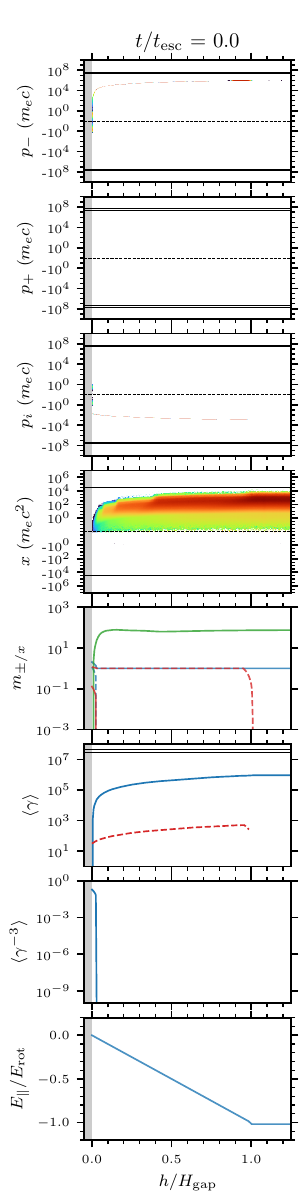}
\includegraphics[clip, trim=1.3cm 0.0cm 0.1cm 0.0cm, height=21cm]{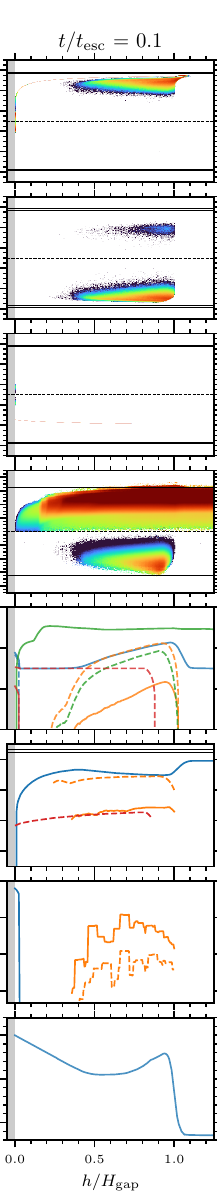}
\includegraphics[clip, trim=1.3cm 0.0cm 0.1cm 0.0cm, height=21cm]{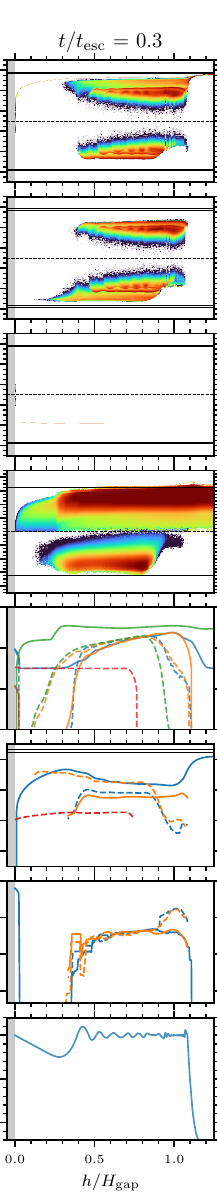}
\includegraphics[clip, trim=1.3cm 0.0cm 0.1cm 0.0cm, height=21cm]{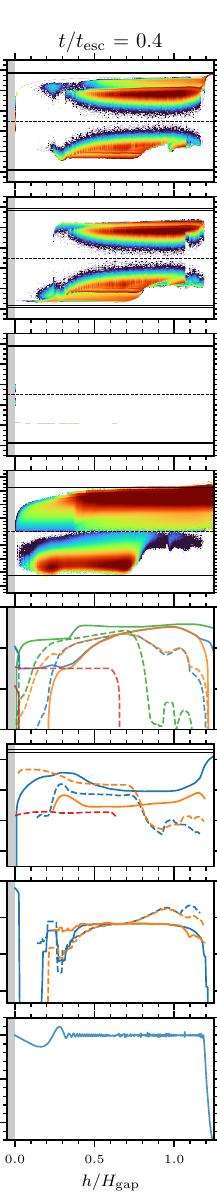}
\caption{\label{fig:gap_sim}
Simulation snapshots of the discharge cycle at $t/t_\mathrm{esc} = 0$ (\textbf{1st column}), $0.1$ (\textbf{2nd column}), $0.3$ (\textbf{3rd column}), and $0.4$ (\textbf{4th column}) for $\alpha = 2$.
\textbf{Rows 1--4:} Particle phase space, $\mathrm{d} n(p)/\mathrm{d} \log p$ (where $p = \gamma\beta$), for electrons (1st row), positrons (2nd), and protons (3rd), and the photon energy spectrum,
$x \,\mathrm{d} n_x(x)/\mathrm{d} \log x$ (4th).
The values range from low to high as blue to green to red.
Vertical thin lines show the $\gamma_\mathrm{gap}$, $\gamma_\mathrm{rad}$, and $x_\mathrm{curv}$ values and the dashed lines the $p = 0$ value.
\textbf{5th row:} Multiplicity of the electrons (blue curve), positrons (orange), and high-energy photons with $x > 2$ (green) for rightwards (solid curves) and leftwards (dashed) moving particles.
\textbf{6th row:} Mean Lorentz factor of electrons (blue) and positrons (orange).
\textbf{7th row:} Inverse plasma-inertia measure, $\langle \gamma^{-3} \rangle$.
\textbf{8th row:} Gap electric field $E_\parallel$ in units of $E_\mathrm{rot}$.
}
\end{figure*}

\begin{figure}[t!]
\centering
\includegraphics[clip, trim=0.0cm 0.0cm 0.0cm 0.0cm, width=8.5cm]{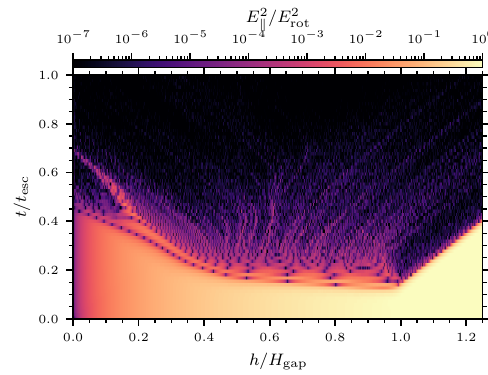}
\caption{\label{fig:discharge_sim_waterfall}
Space-time visualization of the squared electric field $E_\parallel^2/E_\mathrm{rot}^2$ during the first discharge.
The excited oscillations are visible as stripes at $t > 0.2 t_\mathrm{esc}$.
}
\end{figure}

\section{Kinetic Plasma Simulations}\label{sect:sims}

We compare the theoretical analysis to numerical simulations.
We study the radiative plasma dynamics of the magnetospheric circuit with fully kinetic PIC simulations using the \textsc{Runko} framework \citep{nattila2022}%
\footnote{Runko is available at \url{https://github.com/hel-astro-lab/runko};
we use \textsc{v4}~\textsc{kiwi} with commit 86946cc in \textsc{v4-qed} branch.
The simulation setup and analysis scripts are available at \url{https://github.com/hel-astro-lab/polarcaps}.}.
The simulations use realistic system parameters together with exact QED cross sections.

\vspace{1cm}
\subsection{Numerical Setup}

We study the circuit with one-dimensional electrodynamic simulations that evolve the magnetic and electric fields, $\vec{B}(h)$ and $\vec{E}(h)$, as a function of the height $h$.
Our fiducial simulation box has a length of $L_\mathrm{box} \sim 3\times 10^5 \Delta x$, where $\Delta x$ is the grid spacing (1024 tiles with 300 mesh points).
We take $H_\mathrm{gap} = 0.8 L_\mathrm{box}$.
The skin depth is $c/\omega_\mathrm{p,co} \approx 33 \Delta x$ so that $H_\mathrm{gap}/(c/\omega_\mathrm{p,co}) \approx 7400$, similar to real radio pulsars (for which the ratio is $\approx 5000$).
We use a numerical speed of light of $\hat{c} = c \Delta t/\Delta x = 0.45$
and evolve the simulation up to $\sim 20 \,t_\mathrm{esc}$ ($10^7$ steps).
We set two particles per cell per species to match the co-rotation number density.
The simulation follows electrons, positrons, and protons (with a realistic mass ratio $m_p/m_e = 1836$).

We take the initial charge density $\eta = 0$ for $h < H_\mathrm{gap}$ and $\eta = \eta_\mathrm{co}$ for $h \ge H_\mathrm{gap}$;
it is obtained by inserting electrons everywhere in the domain and protons for $h < H_\mathrm{gap}$.
We solve the initial electric field from $\vec{\nabla} \cdot \vec{E} = 4\pi (\eta - \eta_\mathrm{co})$, resulting in $\vec{E}(h)$ with a linear ramp profile between $E_h(0) = 0$ (atmosphere) and $E_h(H_\mathrm{gap}) = E_\mathrm{rot}$ (co-rotating magnetosphere), where $E_h \equiv E_\parallel$ is the field component along $\nvec{h}$.
We take the magnetic field to be constant, $\vec{B}= B_\star \nvec{h}$.

We advance the electromagnetic fields with a second-order field solver.
A first-order method interpolates particle positions and a ZigZag algorithm deposits the current.
We push the particles with a Boris particle pusher.
The current is smoothed with a digital filter (with a binomial 3-point shape) with 8 passes per step.
In addition to the standard PIC algorithm, we add a virtual current to model the system in a co-rotating frame with $\beta_\mathrm{rot} = 3 \times 10^{-6}$, equal to the physical value (see Appendix~\ref{app:corot}).
The twist is imposed by an external current with a fiducial value $j_\mathrm{m} = 2 j_\mathrm{co}$, i.e., $\alpha = 2$.
We have also tested $\alpha = -2$ with results qualitatively similar to those of the fiducial case.

The left simulation boundary includes the star and the atmosphere.
The atmosphere supplies particles to the magnetosphere such that the resulting current will always screen the electric field for $h < 0$.
The atmospheric particles are initialized with a temperature $k_\mathrm{B}T/m_e c^2 = 0.3$; we have verified that the subsequent evolution is insensitive to this choice.
The right simulation boundary models the magnetosphere with a free-flow region for particles.
The escaping EM fields are damped to vacuum values in a perfectly matched layer.

The QED module and the adaptive Monte Carlo routines are discussed in \citet{nattila2024} (Supplementary Material).
Here, we take into account synchrotron radiation and one-photon (nonlinear Breit--Wheeler) pair creation.
We re-scale the QED processes by a factor $\mathcal{N}_\mathrm{mp} = 2 \times 10^{11}$;
i.e., each interacting computational particle is taken to consist of $\mathcal{N}_\mathrm{mp}$ real particles.
This sets the numerical electron radius,
$r_e/\Delta x =  (4\pi \mathcal{N}_\mathrm{mp} \hat{c}^2)^{-1}$, and gives a realistic scale separation of $r_e/H_\mathrm{gap} \sim 10^{-17}$.
A Monte Carlo sub-cycling technique is used to sample synchrotron losses when $\Delta t > t_x$ to ensure that particles cool to the correct value irrespective of the time step.
The field-line curvature is modeled in one dimension by inserting a virtual $B_y$ component into the synchrotron calculation for the particles.
For the photons, we track the accumulated angle w.r.t.\ the field line with $B_y \propto \sin([h-h_0]/R_\mathrm{curv})$, where $h_0$ is the location where the photon was produced.
The QED processes are artificially inhibited in the upper magnetosphere with $h > H_\mathrm{gap}$ to model a finite gap length.
We set $b = 0.02$ and $R_\mathrm{curv} = 30 H_\mathrm{gap}$.
These choices result in $\gamma_\mathrm{rad}/\gamma_\mathrm{gap} \approx 0.3$ and $l_\mathrm{mfp}/H_\mathrm{gap} \approx 0.01$.
The maximum particle energy is $\gamma_\mathrm{gap} \approx 3 \times 10^7$.
In discharge-model units, these choices correspond to $t_\mathrm{acc}/t_\mathrm{esc} \approx 2 \times 10^{-8}$ and $t_\pm/t_\mathrm{esc} = l_\mathrm{mfp}/H_\mathrm{gap} \approx 10^{-2}$.

The simulation setup differs only slightly from previous simulations (notably \citealt{timokhin2010,timokhin2013}) by having realistic system parameters, employing exact differential cross sections for the QED processes, including a thin, gravitationally bound atmosphere, and solving the EM fields in the co-rotating frame.
These improvements result in a more realistic particle distribution during the discharge, which, in turn, give a more realistic inverse-plasma-inertia measure $\langle \gamma^{-3} \rangle$.

\subsection{Observed Dynamics}

\begin{figure}[t!]
\centering
\includegraphics[clip, trim=0.0cm 0.0cm 0.0cm 0.0cm, width=8.5cm]{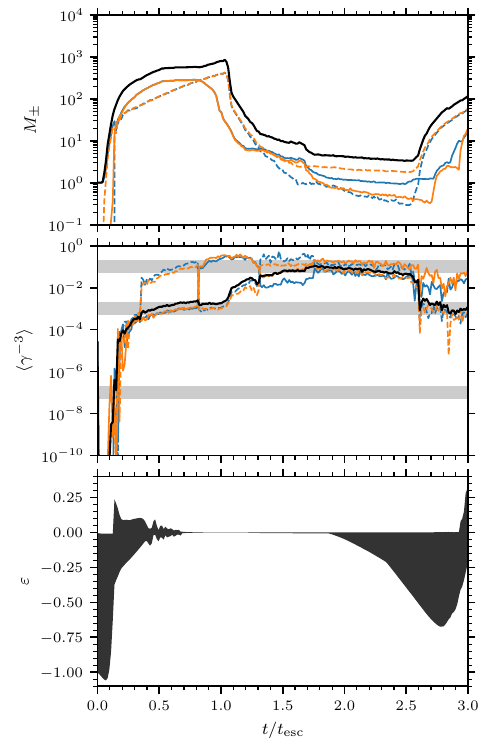}
\caption{\label{fig:discharge_sim}
Measurements of the discharge region quantities from a full kinetic simulation during the first pair cascade event.
Top panel shows the local peak multiplicity $M_\pm$ found in the gap zone (solid black curve).
Multiplicities of electrons (blue) and positrons (orange) are also shown with a split between upwards (solid curves) and downwards (dashed) moving particle populations.
The second panel shows the inverse plasma inertia $\langle \gamma^{-3}\rangle$ measured at the location coinciding with the peak total multiplicity.
Three bands are highlighted as characteristic values:
ignition phase with $\langle \gamma^{-3} \rangle \sim 10^{-7}$, saturation $\sim$$10^{-3}$, and decay $\sim$$0.1$.
Third panel shows the minimum and maximum values of the electric field.
}
\end{figure}

\begin{figure}[t!]
\centering
\includegraphics[clip, trim=0.0cm 0.0cm 0.0cm 0.0cm, width=8.5cm]{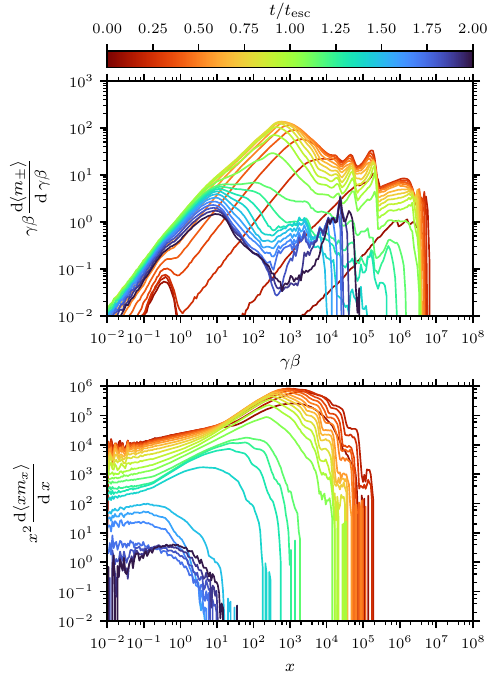}
\caption{\label{fig:discharge_sim_spec}
Average particle spectra in the gap as a function of time.
Top panel shows the average momentum distribution for the pair plasma across the full gap (i.e., $h \in [0, H_\mathrm{gap}]$) and the bottom panel the photon energy spectra.
}
\end{figure}

The particle dynamics in the gap follow the picture described in Sect.~\ref{sect:4}.
The evolution is visualized in Fig.~\ref{fig:gap_sim}.
In the beginning, we observe a rapid acceleration of the particles to $\gamma_\mathrm{rad}$.
They then emit copious synchro-curvature photons at the radiation limit.
The photons travel a distance of $\sim l_\mathrm{mfp}$ during which their pitch angle against the field line increases until the photon undergoes one-photon pair creation.

The subsequent gap dynamics follow the phases outlined in Sect.~\ref{sect:discharge}.
The empty gap is loaded with photons.
The gap stays in the loading and ignition phases until the plasma cloud develops a moderate bulk velocity and an electric current forms.
The current then screens the electric field.
The abrupt screening leads to overshooting the equilibrium value and the electric field in the gap is left ringing with the frequency $\omega_\mathrm{osc}$.
Fig.~\ref{fig:discharge_sim_waterfall} visualizes the $E_h^2(h,t)$ dynamics as a space-time diagram;
the oscillations are clearly visible as ``stripes'' starting at $t \approx 0.2 t_\mathrm{esc}$ and $h \in [0.5, 1] H_\mathrm{gap}$.
At the end of the discharge, the seed particles created during the saturation phase decay into next-generation particles.
The gap region is flushed during $\sim t_\mathrm{esc}$ and the accelerating electric field begins to be restored.
The cycle time is shorter than the analytic estimate given by Eq.~\eqref{eq:t_cycle} because the effective gap length is closer to $\sim 0.2 H_\mathrm{gap}$ in our simulations, augmented by the linear electric field profile.

The global gap dynamics are visualized in Fig.~\ref{fig:discharge_sim} for the first discharge-episode and the results can be compared to the numerical solutions of the concurrency model (see Fig.~\ref{fig:discharge_model}).
The results are in reasonable agreement.
The peak multiplicity evolves broadly as presented in Sect.~\ref{sect:discharge}:
it grows exponentially, saturates, reaches a maximum, and, finally, decays exponentially.
The main difference is the prolonged plateau-like saturation phase, caused by the inclusion of particle populations beyond the second generation.
Another visual difference is in the decay phase: the pair cloud is moving and slides out of the finite simulation domain at around $t \approx 1.1 t_\mathrm{esc}$, causing the peak multiplicity to decay abruptly.

The inverse plasma inertia $\langle \gamma^{-3} \rangle$ is an important system parameter that controls the cascade physics.
Its exact value is determined by the delicate competition between particle acceleration and QED processes (which keep the pairs relativistic and $\langle \gamma^{-3} \rangle$ small) and plasma thermalization (which increases $\langle \gamma^{-3} \rangle$).
During the loading phase, the unscreened electric field accelerates the pairs to high velocities, maintaining a small value of $\langle \gamma^{-3} \rangle_\mathrm{init} \lesssim 10^{-10}$.
At the onset of ignition, when $m_\pm \sim 1$, we measure $\langle \gamma^{-3} \rangle_\mathrm{ign} \sim 10^{-7}$.
During the saturation phase ($t \sim t_\mathrm{esc}$) the distribution broadens and a pronounced peak forms with a mean Lorentz factor $\langle \gamma \rangle \approx \gamma_\mathrm{min} \sim 400$;
the resultant $\langle \gamma^{-3} \rangle_\mathrm{sat} \sim \frac{1}{2} \langle \gamma \rangle^{-1} \sim 10^{-3}$ (see Appendix~\ref{app:g3}).
Late in the decay ($t \sim 2 t_\mathrm{esc}$), we observe an even cooler thermal peak forming with $\langle \gamma\rangle \sim 10$;
it corresponds to $\langle \gamma^{-3} \rangle_\mathrm{late} \sim  0.1$.
We caution that the exact value of $\langle \gamma^{-3}\rangle_\mathrm{late}$ is numerically demanding to resolve and its value might not be fully converged; this, however, does not change our main conclusions.

The general features of the discharges are very similar for the $\alpha < 0$ case \citep{timokhin2010, timokhin2013}.
In particular, we have tested $\alpha = -2$ and found results similar to those of the fiducial case.

\subsection{Peak Multiplicity}

The measured peak multiplicity is $M_\pm \sim 10^3 - 10^4$ in the fiducial simulation;
the gap-averaged value is $\langle M_\pm \rangle \sim \mathrm{a}~\mathrm{few} \times 10^2$.
According to our discharge model, the peak multiplicity is set mainly by the unknown plasma inertia, for which we measure $\langle \gamma^{-3} \rangle_\mathrm{ign} \sim  10^{-7}$ during the cascade ignition.
Using Eq.~\eqref{eq:Mpm} gives an estimate for the peak multiplicity of $M_\pm \sim 5 \times 10^3$.

The measured pair yield varies depending on the simulation input parameters and modeling choices, such as $\gamma_\mathrm{gap}$, $\gamma_{\mathrm{rad}}$, and gap $\vec{E}$-field profile shape.

\subsection{Particle Distributions}

\begin{figure}[t!]
\centering
\includegraphics[clip, trim=0.0cm 0.0cm 0.0cm 0.0cm, width=8.5cm]{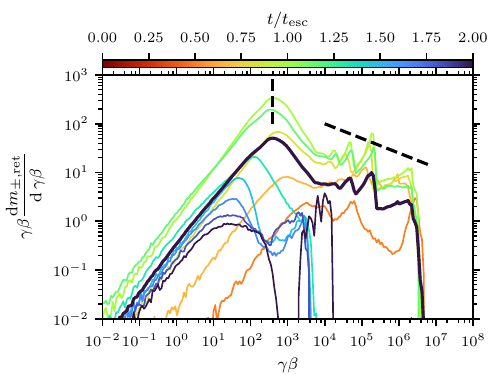}
\caption{\label{fig:discharge_sim_return_spec}
Return-particle spectra as a function of time (colored curves).
The time-averaged spectrum of particles bombarding the atmosphere is shown with a thick black curve.
The vertical dashed line corresponds to $\gamma = 400$ and the tilted line to a power-law slope of $\gamma^{-0.3}$.
}
\end{figure}

Particle spectra during the discharge are visualized in Fig.~\ref{fig:discharge_sim_spec}.
The particles are spatially averaged in a region $h \in [0, 1] H_\mathrm{gap}$.
At early times ($t \lesssim 0.2 t_\mathrm{esc}$), the spectrum peaks strongly at a high Lorentz factor $\gamma_1 \sim 10^7$ ($\approx \gamma_\mathrm{rad}$).
The peak remains slightly below $\gamma_\mathrm{rad}$ because $E_h$ is not exactly $E_\mathrm{rot}$ when the discharge starts.
The photons then decay into secondary pairs, forming a second peak at $\gamma_2 \sim 10^5$ ($\sim 10^{-2} \gamma_\mathrm{rad}$).
In the saturation phase ($t \sim [0.5$--$1] \,t_\mathrm{esc}$), the cascade proceeds to lower and lower energies creating the subsequent generations.
In the spectra, we see the formation of particles up to the sixth generation.

The photon spectra have similar peaks at $x_1 \sim 10^{-2} \, \gamma_1$, $x_2 \sim 10^{-1} \, \gamma_2$, etc., at early times ($t \lesssim 0.5 t_\mathrm{esc}$).
The number density of high-energy gamma rays ($x \sim 10^2$ to $10^5$) drops quickly in the saturation phase as the photons are converted into pairs.
Only low-energy gamma rays ($x \lesssim 10$) exist in the late decay phase ($t \gtrsim t_\mathrm{esc}$).

The particle distributions are not exactly the same for electrons vs.\ positrons and for inward- vs.\ outward-moving populations; instead they evolve as a function of time.
At early times ($t \lesssim 0.2 t_\mathrm{esc}$; loading), the distributions at different heights for outward-moving electrons and inward-moving positrons resemble each other, and vice versa, because the particles respond similarly to the $\vec{E}$ field.
Later ($t \sim [0.3$--$1] \,t_\mathrm{esc}$; saturation), the distributions for electrons and positrons moving in the same direction resemble each other, but differences appear between inward and outward populations (see Fig.~\ref{fig:gap_sim}).
For late times ($t \gtrsim t_\mathrm{esc}$; decay), all the distributions start to resemble each other since the particles have had time to advect and mix.

\subsection{Return-Particle Flux}

The characteristics of the return-particle flux are important because the in-falling particles cause heating of the atmosphere \citep{salmi2020}.
The particle distribution for the in-falling pairs is visualized in Fig.~\ref{fig:discharge_sim_return_spec}.
The distribution closely resembles the overall particle spectra in the gap (see Fig.~\ref{fig:discharge_sim_spec}) because the majority of the particles are produced at the late saturation stage of the burst, during which the particle populations have had time to mix.

A thermal distribution with $\langle \gamma \rangle \sim 400$ and a power-law tail extending up to $\sim 0.1 \gamma_\mathrm{rad}$, $\ud m_\pm / \ud \ln \gamma \propto \gamma^{-s}$ with $s \approx 0.3$, models the time-averaged distribution.

\subsection{Electromagnetic Oscillations}

\begin{figure}[t!]
\centering
\includegraphics[clip, trim=0.0cm 0.0cm 0.0cm 0.0cm, width=8.5cm]{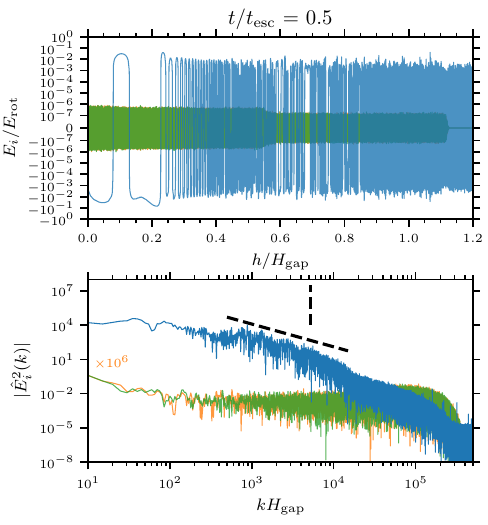}
\caption{\label{fig:discharge_sim_em_spec}
Top panel shows the electric field oscillations in the gap for $E_h$ (blue), $E_y$ (orange), and $E_z$ (green).
Bottom panel shows the Fourier spectra of the electric field for $|\hat{E}_h|^2$, $|\hat{E}_y|^2$, and $|\hat{E}_z|^2$ ($y$ and $z$ components are shifted by $\times10^6$).
Spectra have slopes of $|\hat{E}_h|^2 \propto k^{-2}$ (tilted dashed line), $|\hat{E}_y|^2 \propto k^{0}$, and $|\hat{E}_z|^2 \propto k^{0}$.
The characteristic oscillation wavenumber $k_\mathrm{osc} = \omega_\mathrm{osc}/c$
(computed for $M_\pm = 5 \times 10^2$ and $\langle\gamma^{-3}\rangle = 10^{-3}$ which are measured from the simulation at $t = 0.5 t_\mathrm{esc}$)
is shown by the vertical dashed line;
it corresponds roughly to the maximum oscillation frequency excited by the discharge.
}
\end{figure}

The discharge excites strong electromagnetic oscillations in the gap.
These oscillations excite (super-luminal-branch) O-mode waves in 2D/3D when the discharge cloud front is tilted w.r.t.\ the local magnetic field, i.e., $\partial_y n_\pm \ne 0$ for $\vec{B} = B \nvec{h} \perp \nvec{y}$ \citep{philippov2020}.
Therefore, it is interesting to analyze the properties of these oscillations because they can be connected to the pulsar radio emission.

The oscillation phase of the discharge cycle excites primarily $E_h$ fluctuations with $\delta E_h/E_\mathrm{rot} \equiv \eta_E \sim 10^{-2}$ (see Fig.~\ref{fig:discharge_sim_em_spec}).
The wavenumber ($k$) power spectrum of these oscillations is broad:
$|\hat{E}_h|^2 \propto k^{-2}$, broadly comparable to real pulsar observations, in which the mean flux density scales as $S_\nu \propto \nu^{-1.4}$ with observing frequency $\nu$ \citep{bates2013,kumar2025}.
The measured power spectrum has additional time-variable features (dips and spikes) at values of $k$ matching the analytic estimate for the oscillation frequency, $k_\mathrm{osc} \approx \omega_\mathrm{osc}/c$, supporting the characteristic system frequency.

As the discharge evolves, the frequency changes.
At ignition, the plasma density has only started to grow from the co-rotation value and the plasma is dominated by the freshly accelerated beam, $\langle \gamma^{-3} \rangle_\mathrm{ign} \sim 10^{-7}$; with $m_\pm \sim 10$ this gives $\omega_\mathrm{osc} t_\mathrm{esc} \sim (m_\pm \frac{t_\mathrm{esc}}{t_\mathrm{acc}} \langle \gamma^{-3} \rangle_\mathrm{ign})^{1/2} \sim 5$ ($\omega_\mathrm{osc}/2\pi \sim 1 \,\mathrm{MHz}$).
The frequency then rises steeply towards the saturation phase, where the multiplicity reaches its peak value $M_\pm \sim 10^4$ and the pair plasma distribution has broadened to have $\langle \gamma^{-3} \rangle_\mathrm{sat} \sim 10^{-3}$, yielding $\omega_\mathrm{osc} t_\mathrm{esc} \sim 10^{4}$ ($\omega_\mathrm{osc}/2\pi \sim 3 \,\mathrm{GHz}$).
Finally, as the particles escape the zone ($m_\pm \rightarrow 1$) and only the slowest, mildly relativistic plasma with $\langle \gamma^{-3} \rangle_\mathrm{late} \sim 0.1$ remains, the frequency drops back to $\omega_\mathrm{osc} t_\mathrm{esc} \sim 10^{3}$ ($\omega_\mathrm{osc}/2\pi \sim 400 \,\mathrm{MHz}$).
The emitted frequency is therefore predicted to sweep upwards during the beginning of the discharge and then fall down as the gap empties.

The excited oscillations are strong.
We can quantify their strength with the nonlinearity parameter
\begin{align}\begin{split}
  a &\equiv \frac{e \delta E_h}{m_e c \, \omega_\mathrm{osc}}
    = \eta_E \left( \frac{2 \gamma_\mathrm{gap}}{ m_\pm \langle \gamma^{-3} \rangle } \right)^{1/2} \\
    &\sim 10 \, \eta_{E,-2} \, \frac{ B_{12}^{1/2} }{ P_0 }
      \left( \frac{ m_\pm }{ 10^4 } \right)^{-1/2}
      \left( \frac{ \langle \gamma^{-3} \rangle }{ 10^{-3} } \right)^{-1/2},
\end{split}\end{align}
expressed using the saturation-phase values.
The parameter is $a \sim 10^{4} \, \eta_{E,-2}$ at ignition and $a \sim 100 \, \eta_{E,-2}$ in the late decay stage.
The observed radio luminosity $L_\mathrm{radio} \sim 10^{28} \,\mathrm{erg}\,\sec^{-1}$ gives an independent comparison
\begin{equation}
  a \sim \frac{4 \pi e \sqrt{L_\mathrm{radio}/c} }{m_e c \, \nu_\mathrm{radio} R_\mathrm{pc}}
    \sim 300 \, L_{28}^{1/2} \, P_0^{1/2}
    \left( \frac{\nu_\mathrm{radio}}{1\,\mathrm{GHz}} \right)^{-1} .
\end{equation}
The luminosity-implied relative amplitude,
$\delta E/E_\mathrm{rot} \sim 4\times10^{-2} \, L_{28}^{1/2} P_0^{2} B_{12}^{-1}$,
is comparable to the measured $\eta_E \sim 10^{-2}$.
Both estimates place the oscillations in the strong-wave regime, $a \gg 1$:
the waves force relativistic particle oscillations with quiver momenta $p \sim a \, m_e c$,
so their subsequent propagation and mode conversion are nonlinear.

The rotating frame (see Appendix~\ref{app:corot}) also enables excitation of the $E_y$ component, but
the resulting amplitudes are much smaller, $\delta E_y \sim 10^{-6} \delta E_h$.
These fluctuations can, however, be important as they enable excitation of the X-mode in the plasma.
Their spectrum is flat ($|\hat{E}_y|^2 \propto k^0$).

\section{Summary}

Radio-pulsar polar cap discharges generate the pair plasma that powers coherent radio emission.
We studied these discharges with analytic estimates and numerical simulations to understand the dynamics of the magnetospheric circuit.

Sects.~\ref{sect:2}--\ref{sect:4} reviewed the plasma physics of polar cap dynamics and re-expressed the historical formulas in unified modern notation.
A new concurrency model (Sect.~\ref{sect:discharge}) simultaneously describes the time-evolution of the plasma multiplicity, photon density, and electric field.
The analytical model encapsulates all key features of pair discharge dynamics: limit-cycle behavior, intermittent pair-plasma generation, and excitation of electric field oscillations.
These cycles connect to the microstructures seen in pulsar radio data.

One-dimensional radiative PIC simulations (Sect.~\ref{sect:sims}) validated the analytic theory.
The simulations employ exact QED cross sections and realistic plasma parameters, enabling direct analysis of the generated particle distributions.
The numerical setup builds on that of \citet{timokhin2010,timokhin2013}---adding realistic system parameters, exact QED cross sections, the atmosphere, and the co-rotating frame---and our results agree well with their work.
We characterized the particle spectra of the discharge clouds and the return-particle showers falling back onto the polar caps.
The electric field oscillations produced by the discharge have more realistic properties than in previous simulations because they are induced by realistic particle distributions (i.e., realistic $\langle \gamma^{-3} \rangle$ during the screening).

This work focused on synchrotron-based cascades, although other processes can also contribute to the discharge physics.
Compton-based cascades, relevant for millisecond pulsars, will be reported in Salmi \& N\"attil\"a (in prep.).

\section*{Acknowledgments}
J.N. and T.S. acknowledge useful discussions with A. Timokhin.
The authors used Anthropic Claude (Opus-5 and Fable-5) models for assistance with manuscript editing and equation checking.

This work is supported by an ERC grant (ILLUMINATOR, 101114623) and by the Research Council of Finland Centre of Excellence in Neutron-Star Physics (project 374063).
TS acknowledges funding by the Research Council of Finland grant No.~368807.
The views and opinions expressed are however those of the authors only and do not necessarily reflect those of the European Union or the European Research Council.
Neither the European Union nor the granting authority can be held responsible for them.
The authors thank the Finnish Computing Competence Infrastructure (FCCI) for supporting this project with computational and data storage resources and for maintaining the HILE cluster.

\textit{Software:}
We use \textsc{runko} \citep{nattila2022} for the kinetic simulations.
Data analysis and visualization software used in this study:
\textsc{numpy} \citep{numpy2020},
\textsc{matplotlib} \citep{matplotlib2007},
\textsc{pybind11} \citep{pybind2017},
and
\textsc{scipy} \citep{scipy2020}.

\bibliographystyle{aasjournal}
\bibliography{refs}

\appendix

\section{Electrodynamics in a Moving Frame}\label{app:corot}

The electrodynamics in a moving frame was first discussed by \citet{schiff1939}.
\citet{fawle1977} and \citet{timokhin2010} applied these equations to pulsar magnetospheres.
Here, we present their main equations and perform a scaling analysis of the terms.

Consider a rotating frame with an angular velocity $\vec{\Omega}$.
The corresponding velocity of a magnetic field bundle at a radius $r = R$ is
$\vec{\beta} = \vec{\Omega} \times \vec{R} /c$.
In this co-rotating frame, the observer experiences a co-moving time $t$ and uses Lorentz-contracted spatial derivatives $\vec{\nabla}$.
Likewise, the electric field $\vec{E}$, magnetic field $\vec{B}$, current $\vec{j}$, and charge density $\eta$ are measured by the moving observer.

Maxwell's equations in the rotating frame at $r = R$ follow from a Lorentz boost to the co-moving frame. After simplification:
\begin{align}
  \vec{\nabla} \cdot \vec{E} &= 4\pi (\eta - \eta_\mathrm{vir} )\\
  \vec{\nabla} \cdot \vec{B} &= 0 \\
  \frac{\partial \vec{B}}{\partial t} & = - c \vec{\nabla} \times \vec{E} \\
  \frac{\partial \vec{E}}{\partial t} & =  c  \vec{\nabla} \times \vec{B} - 4\pi (\vec{j} - \vec{j}_\mathrm{vir} ) \, .
\end{align}
The equations appear similar to the standard Maxwell equations except for a
virtual charge density
\begin{align}\begin{split}\label{eq:vir_eta}
  \eta_\mathrm{vir} &=
-\frac{\vec{\Omega} \cdot \vec{B}}{2\pi c} +
\frac{r}{4\pi R} \vec{\beta} \cdot \vec{\nabla} \times \vec{B} \, ,
\end{split}\end{align}
and current
\begin{align}\begin{split}\label{eq:vir_cur}
  \vec{j}_\mathrm{vir} = -\frac{c r}{4\pi R}\Big\{  & \vec{\nabla} \times (\vec{\beta} \times \vec{E} )
                         + \,  \vec{\beta} \times (\vec{\nabla} \times \vec{E} ) \\
                         &- \, \vec{\nabla} \times [ \vec{\beta} \times ( \vec{\beta} \times \vec{B} ) ]
\Big\} \, .
\end{split}\end{align}
Here the first term in Eq.~\eqref{eq:vir_eta} is the co-rotation charge density $\eta_\mathrm{co} = -\vec{\Omega} \cdot \vec{B} /(2\pi c)$.
It can induce a maximum current $j_\mathrm{co} \equiv \eta_\mathrm{co} c$.

Next, we consider a one-dimensional system along $\nvec{h}$ so that
$\vec{B} = \vec{B}(h)$ and
$\vec{E} = \vec{E}(h)$.
We take the velocity of the frame to be perpendicular to the $\nvec{h}$-axis and express
the magnetic and electric fields as
$\vec{B} = B_\parallel \nvec{h} + \vec{B}_\perp$
and
$\vec{E} = E_\parallel \nvec{h} + \vec{E}_\perp$,
respectively.
We also take $|\vec{\nabla}| \sim l_\parallel^{-1}$, where $l_\parallel$ is the characteristic length scale of variations along $\nvec{h}$.
Then, the order-of-magnitude upper bounds for the different terms at radii $r \sim R$ are
\begin{equation}
\frac{|\eta_\mathrm{vir}|}{|\eta_\mathrm{co}|}
\sim \mathcal{O}(1) + \mathcal{O}\left( \frac{B_\perp}{B_\parallel} \frac{R}{l_\parallel} \right)
\end{equation}
and
\begin{equation}
\frac{|j_\mathrm{vir}|}{|j_\mathrm{co}|} \sim
\mathcal{O}\left( \frac{E_\parallel}{B_\parallel} \frac{R }{l_\parallel} \right) +
\mathcal{O}\left( \frac{E_\perp}{B_\parallel} \frac{R}{l_\parallel} \right)+
\mathcal{O}\left(\beta  \frac{B_\perp}{B_\parallel} \frac{R}{l_\parallel} \right)
\, .
\end{equation}
Assuming $B_\perp/B_\parallel \ll l_\parallel/R$, $E_\parallel/B_\parallel \sim \beta$, $\beta \ll l_\parallel/R$, and $E_\perp \sim B_\perp$, we recover $\eta_\mathrm{vir} \approx \eta_\mathrm{co}$ and $|j_\mathrm{vir}| \ll |j_\mathrm{co}|$.

These assumptions do not necessarily hold for the smallest kinetic-scale waves with
$l_{\parallel,\mathrm{min}} \sim c/\omega_\mathrm{p} \ll R$.
In that limit, the correction terms in Eq.~\eqref{eq:vir_cur} scale as $R/l_\parallel$
and may become non-negligible.
In particular, for predominantly electrostatic fluctuations
($E_\parallel \gg E_\perp$, $B_\perp \approx 0$),
the first term,
$\delta \vec{j}_1 \propto \vec{\nabla} \times (\vec{\beta} \times \vec{E})$,
can dominate and generate transverse currents
$\delta \vec{j}_1 \perp \nvec{h}$.

\section{Analysis of the Discharge Model}\label{app:discharge}

Here we present a self-contained dynamical-systems analysis of the coupled ODE system (Eqs.~\eqref{eq:dis1}--\eqref{eq:dis4}).
We define the shorthand $g_3 \equiv \langle \gamma^{-3} \rangle$ in this appendix.

The ordering of timescales given in Eq.~\eqref{eq:timescales} reveals three well-separated regimes among the ODE timescales: $t_\mathrm{acc}$ (fastest), $t_\pm \sim t_x$ (intermediate), and $t_\mathrm{esc}$ (slowest).
This singular perturbation structure justifies the phase-by-phase matched-asymptotic analysis in the main text.

\subsection{Block Structure and Linear Stability}\label{app:fixedpoints}

The system (Eqs.~\eqref{eq:dis1}--\eqref{eq:dis4}) can be expressed in quasi-linear form as $\dot{\vec{x}} = \mathcal{J}(\vec{x})\,\vec{x} + \vec{s}$,
with state vector $\vec{x} = (m_\pm,\, m_x,\, \varepsilon,\, j)^\mathrm{T}$,
source $\vec{s} = (0,\, 0,\, \alpha/t_\mathrm{esc},\, 0)^\mathrm{T}$, and
\begin{equation}\label{eq:ode_matrix}
\mathcal{J} =
\begin{pmatrix}
  -\dfrac{1}{t_\mathrm{esc}} &
  \dfrac{2}{t_\pm} &
  0 &
  0 \\[8pt]
  \dfrac{\mathcal{H}}{t_x} &
  -\dfrac{1}{t_\pm}-\dfrac{1}{t_\mathrm{esc}} &
  0 &
  0 \\[8pt]
  0 &
  0 &
  0 &
  -\dfrac{1}{t_\mathrm{esc}} \\[8pt]
  0 &
  0 &
  \dfrac{g_3 m_\pm}{t_\mathrm{acc}} &
  -\dfrac{1}{t_\mathrm{esc}}
\end{pmatrix} \, .
\end{equation}

The Heaviside gate $\mathcal{H} \equiv \mathcal{H}(|\varepsilon|-1)$ controls the photon source (entry $(2,1)$), while the explicit dependence on $m_\pm$ in entry $(4,3)$ introduces nonlinearity into the field-current subsystem.
For a fixed gate state $\mathcal{H}$, the pair-photon sector (upper-left $2\times2$ block) evolves independently of the field-current subsystem.
We analyze each block separately below.

\textbf{Screened regime ($\mathcal{H} = 0$).}
Setting $\mathcal{H} = 0$ in the upper-left block of Eq.~\eqref{eq:ode_matrix} gives the pair-photon block of the coefficient matrix
\begin{equation}
\mathcal{J}_\mathrm{scr} = \begin{pmatrix} -1/t_\mathrm{esc} & 2/t_\pm \\ 0 & -1/t_\pm - 1/t_\mathrm{esc} \end{pmatrix} \, .
\end{equation}
This upper-triangular matrix has eigenvalues $-1/t_\mathrm{esc}$ and $-(t_\mathrm{esc} + t_\pm)/(t_\mathrm{esc}\, t_\pm)$, both negative.
The field-current block
\begin{equation}
\mathcal{J}_\mathrm{fld} = \begin{pmatrix} 0 & -1/t_\mathrm{esc} \\ g_3 m_\pm / t_\mathrm{acc}  & - 1/t_\mathrm{esc} \end{pmatrix}
\end{equation}
has the eigenvalues
\begin{equation}
-\frac{1}{2t_\mathrm{esc}} \pm \sqrt{ \frac{1}{4t_\mathrm{esc}^2} - \frac{g_3 m_{\pm}}{t_\mathrm{esc} t_\mathrm{acc}} } \, ,
\end{equation}
whose real part is always negative.
These eigenvalues become complex conjugates, corresponding to damped oscillations, if
\begin{equation}
\frac{4 g_3 m_{\pm}}{t_\mathrm{acc}} > \frac{1}{t_\mathrm{esc}} \, .
\end{equation}

Thus, all modes decay exponentially and the screened state acts as a local attractor.
Pair and photon perturbations decay exponentially, while the current relaxes towards $\alpha$.

\textbf{Unscreened regime ($\mathcal{H} = 1$).}
During ignition, when $t\ll t_\mathrm{esc}$, the escape terms can be neglected.
The pair-photon block then becomes
\begin{equation}\label{eq:ignition_matrix}
\mathcal{J}_\mathrm{unscr} = \begin{pmatrix} 0 & 2/t_\pm \\ 1/t_x & -1/t_\pm \end{pmatrix} \, .
\end{equation}
This matrix has a positive eigenvalue $\Gamma_1 > 0$ (derived in Appendix~\ref{app:ignition}), so the unscreened state is unstable.
The pair-creation instability drives exponential growth until the screening current catches up and flips the Heaviside gate.

Together with particle escape and subsequent field recovery, the alternation between cascade growth and screening produces the limit-cycle behavior discussed in the main text.

\subsection{Relaxation Oscillation Structure}\label{app:relaxation}

The discharge cycle has fast phases (ignition and saturation, timescale $\sim t_\pm$) and slow phases (oscillation and decay, timescale $\sim t_\mathrm{esc}$).
The fiducial ratio $t_\pm/t_\mathrm{esc} = 10^{-3}$ classifies the system as strongly relaxational: fast phases produce large excursions in $m_\pm$ (from $\sim 1$ to $\sim M_\pm$), while slow phases govern gradual decay and field restoration.

The pair-photon sector $(m_\pm, m_x)$ and the field-current sector $(\varepsilon, j)$ couple through two mechanisms:
\begin{enumerate}
\item Current generation: the term $m_\pm \varepsilon\, g_3 / t_\mathrm{acc}$ in $\dot{j}$ converts growing pair density into screening current.
\item Heaviside gate: the factor $\mathcal{H}(|\varepsilon| - 1)$ in $\dot{m}_x$ switches photon production on or off.
\end{enumerate}
This coupling is asymmetric in timescale: pairs evolve on $t_\pm$ (fast), the field responds on $\omega_\mathrm{osc}^{-1}$ (intermediate), and current relaxes on $t_\mathrm{esc}$ (slow).
Screening is abrupt because of this mismatch.

During decay, $\dot{\varepsilon} \approx \alpha/t_\mathrm{esc}$, giving $\varepsilon(t) = \alpha\, t/t_\mathrm{esc}$.
The gap reopens when $|\varepsilon| = 1$, i.e., after a reformation time of $t_\mathrm{esc}/\alpha$.
The full cycle duration, dominated by the slow decay phase, is $t_\mathrm{cycle} \approx t_\mathrm{esc} \ln (M_\pm/|\alpha|)$ (Eq.~\eqref{eq:t_cycle}).

\subsection{Ignition Phase: Eigenvalue Analysis and Analytic Solutions}\label{app:ignition}

For $t \ll t_\mathrm{esc}$ and $\mathcal{H} = 1$, escape losses are negligible and the pair-photon system reduces to the $2 \times 2$ ignition matrix $\mathcal{J}_\mathrm{unscr}$ (Eq.~\eqref{eq:ignition_matrix}):
\begin{align}
\dot{m}_\pm &= \frac{2 m_x}{t_\pm} \, , \\
\dot{m}_x   &= -\frac{m_x}{t_\pm} + \frac{m_\pm}{t_x} \, .
\end{align}
The characteristic equation is
\begin{equation}\label{eq:char_ign}
\Gamma^2 + \frac{\Gamma}{t_\pm} - \frac{2}{t_\pm\, t_x} = 0 \, ,
\end{equation}
with eigenvalues
\begin{equation}\label{eq:eigenvalues}
\Gamma_{1,2} = \frac{-1 \pm \sqrt{1 + 8\, t_\pm/t_x}}{2\, t_\pm} \, .
\end{equation}
The positive root gives the growth rate $\Gamma_1 = 1/(2t_\mathrm{ign}) - 1/(2t_\pm)$, where we define the ignition timescale
\begin{equation}
   t_\mathrm{ign} \equiv \frac{t_\pm \sqrt{t_x}}{\sqrt{8\, t_\pm + t_x}} \, .
\end{equation}
For $t_\pm \sim t_x$: $t_\mathrm{ign} \approx t_\pm/3$, $\Gamma_1 = 1/t_\pm$, $\Gamma_2 = -2/t_\pm$.
Solving the eigenvector equation for $\Gamma_1$ gives $m_x/m_\pm = \Gamma_1\, t_\pm/2$.
For $t_\pm = t_x$, this ratio equals $1/2$: during ignition, photons are roughly half as numerous as pairs.

With initial conditions $m_\pm(0) = m_{\pm,0}$ and $m_x(0) = 0$,
the eigenmode decomposition gives
\begin{equation}
\begin{split}
m_\pm(t) = & \, m_{\pm,0}  \exp\left( -\frac{t}{2t_\pm} -\frac{t}{2t_\mathrm{ign}} \right)  \\ & \times \left\{ \frac{t_\mathrm{ign}}{2 t_\pm} \left[
\exp\left(\frac{t}{t_\mathrm{ign}} \right)
- 1\right]
+  \frac{1}{2}  \left[
\exp\left(\frac{t}{t_\mathrm{ign}} \right)
+ 1\right] \right\} \, ,
\end{split}
\end{equation}
and
\begin{equation}
m_x(t) = m_{\pm,0}\frac{ t_\mathrm{ign}}{t_x} \exp\left( -\frac{t}{2t_\pm} -\frac{t}{2t_\mathrm{ign}} \right) \left[
\exp\left(\frac{t}{t_\mathrm{ign}} \right)
- 1\right] \, .
\end{equation}
For $t \gtrsim t_\mathrm{ign}$ (or equivalently $t \gtrsim t_\pm$ when $t_\mathrm{ign}\sim t_\pm$), the growing mode dominates and both
components grow with the same exponential rate,
\begin{equation}
m_\pm(t) \propto m_x(t) \propto
\exp\left( \frac{t}{2t_\mathrm{ign}} - \frac{t}{2t_\pm} \right) \, .
\end{equation}
For $t_\pm=t_x$, this growth factor becomes $\exp(t/t_\pm)$.

The full ignition matrix including escape is
\begin{equation}
\mathcal{J}_\mathrm{unscr}^\mathrm{full} = \begin{pmatrix} -1/t_\mathrm{esc} & 2/t_\pm \\ 1/t_x & -1/t_\pm - 1/t_\mathrm{esc} \end{pmatrix} \, .
\end{equation}
The eigenvalues shift by $-1/t_\mathrm{esc}$ relative to $\mathcal{J}_\mathrm{unscr}$.
For the fiducial parameters, the correction to $\Gamma_1$ is $\approx 0.1\%$.

Growth requires the shifted growth eigenvalue to remain positive, equivalently $\det(\mathcal{J}_\mathrm{unscr}^\mathrm{full}) < 0$.
This condition reduces to
\begin{equation}\label{eq:bifurcation}
t_x < t_{x,\mathrm{crit}} = \frac{2\, t_\mathrm{esc}^2}{t_\mathrm{esc} + t_\pm} \approx 2\, t_\mathrm{esc} \, ,
\end{equation}
a bifurcation on the Heaviside switching boundary.
The fiducial value $t_x/t_\mathrm{esc}  \ll 2$ places the discharge far above the ignition threshold.

\subsection{Saturation and Decay Phases}\label{app:saturation}

\textbf{Saturation.}
When $\mathcal{H} \to 0$ (screened), the pair-photon system becomes
\begin{align}
\dot{m}_\pm &= 2\frac{m_x}{t_\pm}  -\frac{m_\pm}{t_\mathrm{esc}} \, , \\
\dot{m}_x   &= -\frac{m_x}{t_\pm} -\frac{m_x}{t_\mathrm{esc}} \, .
\end{align}
The photon solution is
\begin{equation}\label{eq:mx_sat_app}
m_x(t) = M_x' \exp\left( -\frac{t}{t_\mathrm{esc}} - \frac{t}{t_\pm}\right) \, ,
\end{equation}
and the pair solution is
\begin{equation}\label{eq:mp_sat_app}
m_\pm(t) = m_x(t)
\left\{
\exp\left(\frac{t}{t_\pm} \right) \left[ \frac{ M_\pm' }{ M_x' } + 2 \right] - 2
\right\} \, ,
\end{equation}
where $M_x'$ and $M_\pm'$ are the multiplicities at the onset of screening.
The peak of $m_\pm(t)$ occurs after a delay $t_\mathrm{d} = t_\pm \ln \{ 2 M_x' (t_\mathrm{esc} + t_\pm)/[ t_\pm (M_\pm' + 2 M_x')] \}$ if
$2 M_x' / M_\pm'  > t_\pm / t_\mathrm{esc}$;
otherwise the maximum occurs at the onset of screening.

\textbf{Decay.}
At late times, $t\gg t_\pm$, the high-energy photon population has decayed, and Eq.~\eqref{eq:mp_sat_app} reduces to
\begin{equation}
m_\pm(t) \approx M_\pm \exp\left(-\frac{t}{t_\mathrm{esc}}\right) \, ,
\end{equation}
where the peak multiplicity is $M_\pm \approx  M_\pm' + 2M_x' \approx 2 M_\pm'$ when $t_x \sim t_\pm$.

\textbf{Current solution.}
The electric field oscillates on timescale $\omega_\mathrm{osc}^{-1} \ll t_\mathrm{esc}$, so the current responds only to the cycle-averaged $\varepsilon$.
The current solution is
\begin{equation}
j(t) = e^{-t/t_\mathrm{esc}} \left( j_\mathrm{d,0} +  M_\pm \varepsilon g_3 \frac{t}{t_\mathrm{acc}}   \right) \, ,
\end{equation}
where $j_\mathrm{d,0} \equiv j(t = 0)$ is the current at the beginning of the decay phase.
At $t = t_\mathrm{esc}$ and assuming $j_{\mathrm{d},0}$ is small,
\begin{equation}
j(t_\mathrm{esc}) \approx m_\pm(t_\mathrm{esc}) \, \varepsilon \frac{ g_3 t_\mathrm{esc}}{t_\mathrm{acc}} \, .
\end{equation}

\subsection{Field-Current Oscillator}\label{app:oscillator}

During the oscillation and decay phases, $m_\pm$ can be treated as quasi-constant over an oscillation period.
Eliminating $j$ from the $(\varepsilon, j)$ equations yields a damped harmonic oscillator,
\begin{equation}\label{eq:oscillator}
\ddot{\varepsilon} + \frac{\dot{\varepsilon}}{t_\mathrm{esc}} + \omega_\mathrm{osc}^2\, \varepsilon = \frac{\alpha}{t_\mathrm{esc}^2} \, ,
\end{equation}
with natural frequency, damping ratio, and quality factor
\begin{equation}\label{eq:osc_params}
\omega_\mathrm{osc}(t) = \sqrt{\frac{ g_3 \, m_\pm(t)}{t_\mathrm{acc}\, t_\mathrm{esc}}} \, , \quad
\zeta = \frac{1}{2\,\omega_\mathrm{osc}\, t_\mathrm{esc}} \, , \quad
\mathcal{Q} = \omega_\mathrm{osc}\, t_\mathrm{esc} \, .
\end{equation}
The amplitude damping time is $t_\mathrm{damp} = 2\,t_\mathrm{esc}$, independent of $m_\pm$ and $g_3$.
The equilibrium field $\alpha\, t_\mathrm{acc}/(m_\pm(t)\, g_3\, t_\mathrm{esc}) \ll 1$ lies deep in the screened regime for sufficiently large pair multiplicities.

\textbf{Underdamped condition.}
The system oscillates ($\zeta < 1$) when $m_\pm > t_\mathrm{acc}/(4\,g_3\, t_\mathrm{esc})$.
For fiducial parameters ($g_3 = \gamma_\mathrm{min}^{-3} \sim 10^{-7}$), the threshold is $m_\pm \sim 0.1$---satisfied for any realistic discharge.
At peak multiplicity, $\mathcal{Q} \sim 100$: the system is strongly underdamped.

\textbf{Frequency chirp during decay.}
For a constant $g_3$, as $m_\pm(t) \propto e^{-t/t_\mathrm{esc}}$ (Eq.~\eqref{eq:m_ph5}), the oscillation frequency chirps downward:
\begin{equation}
\omega_\mathrm{osc}(t) \propto \sqrt{m_\pm(t)} \propto e^{-t/(2\,t_\mathrm{esc})} \, .
\end{equation}
When the concurrent growth of $g_3$ during thermalization is included, however, the increase of $g_3$ outweighs the decay of $m_\pm$ and the net frequency drift over the full cycle is upward.
If we include $\langle \gamma^{-3} \rangle$ dynamics to the ODE group, we find that the frequency first sweeps up and then downwards.

\subsection{Peak Multiplicity}\label{app:currentlag}

We derive an estimate for $M_\pm$ from the condition that the electric field returns to the critical value $\varepsilon = 1$ during ignition.

\textbf{Quasi-static current.}
Setting $\dot{j} = 0$ with $\varepsilon = 1$ in Eq.~\eqref{eq:dis4} defines the quasi-static current
\begin{equation}\label{eq:jqs}
  j_\mathrm{qs} \equiv \frac{m_\pm\, g_3\, t_\mathrm{esc}}{t_\mathrm{acc}} \, .
\end{equation}

\textbf{Current lag.}
During ignition, $m_\pm(t) = m_{\pm,0}\, e^{\Gamma_1 t}$ and $\varepsilon \approx 1$.
Using Eq.~\eqref{eq:jqs}, the current equation becomes $\dot{j} = -(j - j_\mathrm{qs})/t_\mathrm{esc}$: the current relaxes toward the instantaneous quasi-static value on a timescale $t_\mathrm{esc}$.
The particular solution $j_\mathrm{p}(t) = C\, e^{\Gamma_1 t}$ has
\begin{equation}\label{eq:current_lag_C}
  C = \frac{j_{\mathrm{qs},0}}{\Gamma_1\, t_\mathrm{esc} + 1} \, ,
\end{equation}
where $j_{\mathrm{qs},0} = m_{\pm,0}\, g_3\, t_\mathrm{esc}/t_\mathrm{acc}$ is the quasi-static current at ignition onset.
The ratio of the actual to the quasi-static current is
\begin{equation}\label{eq:jp_jqs}
  \frac{j_\mathrm{p}}{j_\mathrm{qs}} = \frac{C}{j_{\mathrm{qs},0}} = \frac{1}{\Gamma_1\, t_\mathrm{esc} + 1} \approx \frac{t_\pm}{t_\mathrm{esc}} \ll 1 \, .
\end{equation}
Thus, the current remains suppressed relative to $j_\mathrm{qs}$ during ignition.

\textbf{Screening from the $\varepsilon$-return condition.}
At the onset of ignition, the current has relaxed to $j_0 < \alpha$ during gap reformation.
Since $j < \alpha$, the field $\varepsilon$ initially rises above~$1$.
The field evolution $\dot{\varepsilon} = (\alpha - j)/t_\mathrm{esc}$, integrated over the ignition phase, gives
\begin{equation}\label{eq:eps_ignition}
  \varepsilon(t) \approx 1 + \frac{(\alpha - j_0)\, t}{t_\mathrm{esc}} - \frac{C\, (e^{\Gamma_1 t} - 1)}{\Gamma_1\, t_\mathrm{esc}} \, ,
\end{equation}
where the first correction is a linear rise (current deficit) and the second is the exponentially growing screening term.
Pair creation ceases when $\varepsilon$ drops below $1$, i.e., $\mathcal{H}(|\varepsilon|-1) \to 0$.
Setting $\varepsilon(t^\ast) = 1$ and using $t^\ast = \ln M_\pm' / \Gamma_1$ (from $M_\pm' = m_{\pm,0}\, e^{\Gamma_1 t^\ast}$ and assuming $m_{\pm,0} = 1$) yields
\begin{equation}\label{eq:Mpm_implicit}
  M_\pm'  = M_0 \ln M_\pm' + 1 \, , \quad
  M_0 \equiv \frac{(\alpha - j_0)\, \Gamma_1\, t_\mathrm{acc}}{g_3} \, .
\end{equation}
The current deficit is $\alpha - j_0 = \mathcal{O}(1)$, giving $M_0 \approx \Gamma_1\, t_\mathrm{acc}/g_3$, nearly independent of $\alpha$.
Since $M_\pm \sim 10^3$--$10^4$ for typical parameters, $\ln M_\pm' \approx 7$--$9$ and the implicit equation reduces to $M_\pm \approx 2 M_\pm' \sim 10\, M_0$.
For $t_\pm = t_x$ (where $\Gamma_1 = 1/t_\pm$), this gives
\begin{equation}\label{eq:Mpm_equal}
  M_{\pm} \sim 10 \frac{t_\mathrm{acc}}{t_x \, g_3} \, .
\end{equation}
Note that the $g_3$ entering the screening current here refers to the plasma population which carry the current during the ignition (i.e., $\langle \gamma^{-3} \rangle_\mathrm{ign}$), not to the loading-phase primary particles (for which $\langle \gamma^{-3} \rangle_\mathrm{init} \sim \gamma_\mathrm{rad}^{-3}$) or the later saturation-stage thermalized (i.e., $\langle \gamma^{-3} \rangle_\mathrm{scr}$).

\textbf{Empirical fitting formula.}
A systematic scan of 36 ODE solutions across five parameter sweeps ($g_3$, $t_\pm$, $t_x$, $t_\mathrm{acc}$, $\alpha$) confirms that the peak multiplicity is predominantly set by $t_\mathrm{acc}/(g_3\, t_x)$, with minimal residual dependence on $t_\pm$ or $\Gamma_1$.
This insensitivity to $t_\pm$ follows from a cancellation: the screening-onset value scales as $\propto \Gamma_1$, while the post-screening yield boost $1 + 2m_x/m_\pm = 1 + \Gamma_1 t_\pm$ contributes the remaining factor, so the peak scales as $\propto \Gamma_1(1 + \Gamma_1 t_\pm) \approx 2/t_x$.
A log-space least-squares fit yields
\begin{equation}\label{eq:Mpm_empirical}
  M_{\pm} \approx 11 \frac{t_\mathrm{acc}}{t_x \, g_3}\, \alpha^{1/4} \, ,
\end{equation}
which tracks the ODE solutions to $\lesssim 30\%$ across all parameter combinations.

\section{On the Estimation of $\langle \gamma^{-3} \rangle$}\label{app:g3}

Approximating the moment $\langle \gamma^{-3} \rangle$ as $\langle \gamma \rangle^{-3}$ is inaccurate for broad particle distributions.
Taylor expansion around the mean $\bar{\gamma} = \langle \gamma \rangle$, with $\delta\gamma \equiv \gamma - \bar{\gamma}$ (see also \citealt{okawa2024}),
\begin{align}\begin{split}
  \langle \gamma^{-3} \rangle &= \left\langle (\bar{\gamma} + \delta \gamma)^{-3} \right\rangle \\ 
                              &= \bar{\gamma}^{-3} \left( 1 + 6 \frac{\langle \delta \gamma^2 \rangle}{\bar{\gamma}^2} - 10
\frac{\langle \delta \gamma^3 \rangle}{\bar{\gamma}^3} + \dots \right)
\end{split}\end{align}
shows that large relative variance in relativistic plasmas renders the series slow to converge.
Jensen's inequality further guarantees that $\langle \gamma^{-3} \rangle \ge \langle \gamma \rangle^{-3}$.
For example, a 1D ultra-relativistic thermal distribution $f(\gamma) \approx \Theta^{-1} e^{-\gamma/\Theta}$ yields $\langle \gamma^{-3} \rangle \approx (2\Theta)^{-1}$ versus $\langle \gamma \rangle^{-3} \approx \Theta^{-3}$.
The ratio $\langle \gamma^{-3} \rangle / \langle \gamma \rangle^{-3} \approx \Theta^2/2$ reaches $\sim 8 \times 10^4$ for our simulated $\Theta \approx \langle \gamma \rangle \sim 400$ when thermalization has occurred.
A smoothly broken power-law model that also captures the high-energy nonthermal tail, $f(u) \propto (u/\Theta)^{\alpha_1} \left[ 1 + (u/\Theta)^\Delta \right]^{(\alpha_2 - \alpha_1)/\Delta}$ (where $\alpha_1 \approx 0$ and $\alpha_2 \approx -1.3$ are the powerlaw indices, and $\Delta \approx 0.3$ is the smoothing factor), gives $\langle \gamma^{-3} \rangle \approx 0.57 \, \Theta^{-1}$.
Robust estimation requires analyzing the discharge stages from numerical simulations.

\end{document}